\documentclass[preprint2]{aastex}

\newcommand{\msun} {{\rm M}_\odot}
\newcommand{\rsun} {{\rm R}_\odot}
\newcommand{\lsun} {{\rm L}_\odot}
\newcommand{\muHz} {\mu\mbox{Hz}}
\newcommand{\Hp} {H_{\rm p}}
\newcommand{\mHz} {{\rm m}{\rm Hz}}
\newcommand{\kms} {\mbox{kms}^{-1}}
\newcommand{\vsini} {v\sin\,i}
\newcommand{\vrot} {v_{{\rm rot}}}

\newcommand{\Os} {\Omega_{\rm s}}
\newcommand{\Oc} {\Omega_{\rm c}}
\newcommand{\kelvin} {{\rm K}}
\newcommand{\teff} {T_{{\rm eff}}}

\newcommand{\ttwo} {\times10^{-2}}
\newcommand{\tthree} {\times10^{-3}}

\newcommand{\etaboo} {$\eta~{\rm Bootis}$}

\newcommand{\bbv} {Brunt-V\"ais\"al\"a}

\newcommand{\lsep} {\Delta\nu_{n,\ell}}
\newcommand{\ssep} {\delta\nu_{n,\ell}}
\newcommand{\alsep} {\langle \Delta\nu_{n,\ell} \rangle}
\newcommand{\assep} {\langle \delta\nu_{n,\ell} \rangle}

\newcommand{\dnu} {\delta\nu}

\newcommand{\dzeroone} {\delta_{0,1}(n)}
\newcommand{\donezero} {\delta_{1,0}(n)}
\newcommand{\rzeroone} {r_{0,1}(n)}
\newcommand{\ronezero} {r_{1,0}(n)}

\newcommand{\nunl} {\nu_{n,\ell}}
\newcommand{\numax} {\nu_{{\rm max}}}
\newcommand{\nuac} {\nu_{{\rm ac}}}
\newcommand{\numaxsun } {\nu_{{\rm max},\odot}}
\newcommand{\nuacsun } {\nu_{{\rm ac},\odot}}

\newcommand{\mumax} {\mu(\numax)}

\newcommand{\Uim} {{\cal U}_i}
\newcommand{\Sim} {{\cal S}_i}

\newcommand{\cesam} {{\sc cesam}}
\newcommand{\filou} {{\sc filou}}
\newcommand{\most} {\emph{MOST}}
\newcommand{\corot} {\emph{CoRoT}}
\newcommand{\kepler} {\emph{Kepler}}

\newcommand{\eqn} [1] {
\begin{equation}
#1
\end{equation}}

\newcommand{\eqna} [1] {
\begin{eqnarray}
#1
\end{eqnarray}}

\shorttitle{The effect of rotation on solar-like oscillations}
\shortauthors{Su\'arez et al.}

\begin{document}

\title{On the interpretation of echelle diagrams for solar-like oscillations \\
       Effect of centrifugal distortion}

\author{J.C. Su\'arez\altaffilmark{1,2} and M.J. Goupil\altaffilmark{2}
        and D.R. Reese\altaffilmark{2} and R. Samadi\altaffilmark{2} 
        and F. Ligni\`eres\altaffilmark{3,4} and M. Rieutord\altaffilmark{3,4} 
        and J. Lochard\altaffilmark{2}}
	
\altaffiltext{1}{Instituto de Astrof\'{\i}sica de Andaluc\'{\i}a (CSIC),
                Granada, Spain.\email{jcsuarez@iaa.es}}
\altaffiltext{2}{Observatoire de Paris, LESIA, CNRS UMR 8109, 92195 Meudon, France.}
\altaffiltext{3}{Universit{\'e} de Toulouse, UPS, Laboratoire d'Astrophysique 
                 de Toulouse-Tarbe (LATT), 31400 Toulouse, France.}                
\altaffiltext{4}{CNRS, Laboratoire d'Astrophysique de Toulouse-Tarbes (LATT), 
                 31400 Toulouse, France.}

\begin{abstract}
    This work  aims at determining the impact of slow to moderate rotation on the regular patterns
    often present in solar-like oscillation spectra, i.e. the frequency spacings. We focus on the
    well-known asteroseismic diagnostic echelle diagrams, examining how rotation may modify the 
    estimates of the large and small spacings, as well as the identification of modes. We illustrate
    the work with a real case: the solar-like star \etaboo.

    We study a main sequence $1.3 M_\odot$ star as a typical case. The modeling takes into account 
    rotation effects on the equilibrium models through an effective gravity and on the
    oscillation frequencies through both perturbative and non-perturbative calculations. We compare
    the results of both type of calculations in the context of the regular spacings (like the small
    spacings and the scaled small spacings) and echelle diagrams. We show that for  echelle diagrams
    the perturbative approach remains valid for rotational velocities up to $40-50\,\kms$.

    We show that for the rotational velocities measured in solar-like stars, i.e.
    $vsini$ up to 20--30 $\kms$, rotation effects must be taken into account in the modeling 
    for a correct interpretation of the observed oscillations. In particular, theoretical
    oscillation frequencies must be corrected up to the second-order in terms of rotation rate,
    including near degeneracy effects. For rotational velocities of about $16\,\kms$ and higher,
    diagnostics on large spacings and on modal identification through echelle diagrams can be
    significantly altered by the presence of the $m\neq0$ components of the rotationally split
    modes. We found these effects to be detectable in the observed frequency range. Analysis of the 
    effects of rotation on small spacings and scaled small spacings reveals that these can be of
    the order of, or even larger than surface effects, typically turbulence, microscopic   
    diffusion, etc.  Furthermore, we show that scaled spacings are significantly affected by
    stellar distortion even for small stellar rotational velocities (from 10-$15\,\kms$) and
    therefore some care must be taken when using them as indicators for probing deep stellar
    interiors.

\end{abstract}

\keywords{stars:~evolution --stars:~individual($\eta$ Bootis) ---
          stars:~interiors -- stars:~oscillations (including pulsations)
          stars:~rotation --stars:~solar-type}

\section{Introduction\label{sec:intro}}
Solar-like oscillations are detected in stars with outer convective zones,  that is, in  
stars similar to the Sun \citep[see][ and references therein]{Michel08sci} as well as in red giant
stars \citep{DeRidder09nat} and even in massive B-type stars as recently shown by
\citet{Kevin09sci}. 
These oscillations are due to intrinsically stable pressure (p) modes which are  excited by
turbulent motions in the convective outer layers. This stochastic process excites several modes in
a star over a broad frequency range well above the fundamental radial mode frequency in a regime
where frequencies show regular patterns in the power spectrum. This gives access to an  information
content based on different seismic signatures which may constrain fundamental stellar properties
(namely mass, age, and radius) and probe the internal stellar structure to unprecedented accuracy
 \citep[see, for instance, ][]{Gough87,Cunha07, Tex09} . 

On the other hand, the uninterrupted long observational runs required together with the expected
small oscillation amplitudes has made the detection of solar-type stars 
technically challenging targets  in particular, for asteroseismic space missions. A wealth of data
is already available from \most\ \citep{most98}, \corot\
\citep[][]{corot03,Michel08sci,Appourchaux08, Kevin09sci} and \kepler\
\citep{Kepler10aip}.

Low mass stars are generally  slow rotators (in most cases $\langle\vsini\rangle<20\,\kms$)
and the influence of rotation on the oscillation frequencies is globally small.
For slow rotators, the geometrical structure is represented by a single spherical harmonics
$Y_\ell^m (\theta, \phi)$, and the frequencies by $\nu_{n,\ell,m}$ with $n$ being the radial order
of the mode. Rotation lifts the degeneracy causing the modes to split into $m\neq0$ multiplets.  The
$m\neq0$ are usually computed using a first-order perturbation formalism. Rotation also
affects, indirectly, the $m=0$ mode frequencies as it modifies the internal structure through
rotationally induced mixing \citep{GoupilTalon02,Mathis07}. In the same line,
\citet{Carrier05etaboo} studied the solar-like star \etaboo\ assuming a surface rotational velocity
of about $\vrot=3-7\,\kms$. The authors found that the effects of including rotationally-induced
mixing was small and the rotation is not strong enough to impose a rigid rotation. In particular,
they found that the core in this star rotates three times faster than
the outer layers.

Distortion due to the centrifugal force can have a much larger impact on the oscillation frequencies
even for slow rotators 
\citep[see][ for a review on rotation effects on the oscillation frequencies]{Goupil09lnp}.
Such an effect is stronger for p modes with small inertia. These modes propagate mainly through the
outer layers of the star. Therefore, their frequencies are more sensitive to changes in the surface
physical properties, where the centrifugal force becomes more efficient. The present work examines
and evaluates the direct effects of rotation on the frequencies in this high-frequency domain, and
its impact on the different seismic diagnostic techniques based on the asymptotic properties of the
oscillations. More specifically, we analyze the case of different modes with close frequencies, such
as $\ell=0$ and $\ell=2$, which are systematically near degenerate. Since the small spacings are
affected by near degeneracy, echelle diagrams, often used as a seismic diagnostic technique for
solar-like stars, should also be affected.

We then first analyzed the effects of rotation on regular patterns observed in solar-like
spectra, focusing on the small and large spacings, which are the most relevant quantities for
echelle spectrograms. Even if solar-like stars are often slow rotators, their oscillation
spectra show high-order frequencies, where the validity of perturbation theory for computing
oscillation frequencies in absence of rotation might fail. This point was thus checked by comparing
the perturbative calculations with recent non-perturbative calculations where the effect of rotation
on the oscillations is fully taken into account \citep{Reese06}. Taking all these previous steps
into account, we investigated  the effect of rotation on echelle diagrams.

The paper is organized as follows: in Section~\ref{sec:sdiag} for sake of notation, we briefly
recall 
the frequency spacings and seismic diagnostics generally used for asteroseismic analysis of
solar-like stars which will be considered in the following sections. Then, the equilibrium stellar 
models and their corresponding oscillation computations are described in Section~\ref{sec:modeling}.
The validity of a perturbative modeling of the effects of rotation on oscillations in solar-like
stars is discussed in Section~\ref{sec:comppoly}. Section~\ref{sec:ED_sm} describes the effects of
rotation for seismic diagnostics based on echelle diagrams. A discussion of the results, including
an illustration based on the solar-like star \etaboo\, and conclusions are given in
Section~\ref{sec:discussion}.

\section{Seismic diagnostics for internal structure\label{sec:sdiag}} 

In the case of solar-like oscillations,modes with  high radial order $n$ are detected. In this
regime, regular patterns are observed in the power spectrum which are related to internal properties
of the star. Analysis of such characteristics allows us to infer information about various physical
processes, particularly those that modify the sound speed and/or the \bbv\ frequency. Several 
frequency combinations are then used to analyze high-order p modes: large and small frequency
spacings and their corresponding scaled counterparts, second-order frequency differences, and even
high-order frequency differences 
\citep[for a review on this topic see][ and references therein]{GoupilDupret07}.

\subsection{Frequency spacings\label{ssec:fspacings}}

The large (frequency) separation is hereafter denoted by $\lsep$ and defined by
\eqn{\lsep=\nu_{n,\ell}-\nu_{n-1,\ell}.\label{eq:def_lsep}}
The large spacing is sensitive to surface layers where $c_s$ is small, and hence oscillation modes
spend more time there compared to the inner regions of the star. This spacing is generally studied 
in order to detect sharp variations of the sound speed, such as those at the base of the solar
convection zone, or those resulting from the presence of ionization regions. The mean large spacing
over several modes in the asymptotic regime scales with the dynamical time scale $(R^3/GM)^{1/2}$ 
and is one of the pieces of information sought when constructing echelle diagrams. 

The small (frequency) spacing, hereafter denoted by $\ssep$, is defined as follows
\eqn{\ssep=\nu_{n-1,\ell+2}-\nu_{n,\ell}.\label{eq:def_ssep}}
This spacing is sensitive to evolutionary effects (the age of the star for a given metallicity).
Average large $\alsep$ and small $\assep$ spacings (over radial order  $n$) are used to estimate
masses and ages of solar-like stars as first suggested by  \citet{JCD98}.

In addition to the aforementioned frequency spacings, other frequency combinations are also used
as efficient asteroseismic diagnostics:
\eqna{\dzeroone & = & \nu_{n,0} - \frac{1}{2}(\nu_{n-1,1} + \nu_{n,1})\\
      \donezero & = & \frac{1}{2}(\nu_{n,0} + \nu_{n+1,0}) - \nu_{n,1},
      \label{eq:def_d01d10}}
together with their corresponding scaled expressions
\eqn{\rzeroone  = \frac{\dzeroone}{\Delta\nu_{n+1,0}},~~~~~
      \ronezero = \frac{\donezero}{\Delta\nu_{n,1}},
     \label{eq:def_r01r10}}
which are sensitive to inner stellar zones \citep[see]{RoxVor01,MazAntia01,RoxVor03}.

\begin{deluxetable}{lcccccccccccc}

  \tablecolumns{13}
  \tablewidth{0pt}
  \tabletypesize{\small}
  \tablecaption{Characteristics of the $1.30\,\msun$ models considered in this work (with the
           exception of ${\cal M}_{\eta}$ whose mass is $1.70\,\msun$).  
           ${\cal P}_i$, and ${\cal P}_i^{\dag}$
           are 1D and 2D polytropic models, respectively. Stellar models are denoted by $\Uim$ and
           $\Sim$, for uniform (constant rotation profile at given evolutionary stage) and
           shellular rotation (rotation profile given by Eq.~\ref{eq:defOmega}) models,
           respectively (details in Section~\ref{ssec:oscill}). Model subscripts $i$
           make reference to the round value of the surface rotational velocity in $\kms$ (for
           ${\cal M}_{\eta}$ the rotational velocity is $30\,\kms$ approximately). The radius of all
           the models is $R=1.276\pm0.005\,\rsun$. Columns represent respectively, from left to
           right, the average large spacing, the acoustic frequency $\nuac$ (Eq.~\ref{eq:defnuac}),
           the value of $\numax$(Eq.~\ref{eq:nu_max}), the rotation frequencies in both the surface
           and the center of the model, and the small numbers $\epsilon$ and $\mu$
           (Eq.~\ref{eq:defeps_mu}) the latter being calculated at the $\numax$ frequency of
           each model.}  
   \tablehead{ 
     \colhead{Model}   & \colhead{$\alsep$} & \colhead{$\nuac$}    & \colhead{$\numax$} &
     \colhead{$\Os$}   & \colhead{$\Oc$}    & \colhead{$\epsilon$} & \colhead{$\mumax$} \\
     \colhead{}        & \colhead{$\muHz$}  & \colhead{$\mHz$}     & \colhead{$\mHz$}   &
     \colhead{$\muHz$} & \colhead{$\muHz$}  & \colhead{$\ttwo$}    & \colhead{$\tthree$}}
   \startdata
    ${\cal P}_{0}^{\dag}$  & 98.40  & -    & -    &    0 &    0 &    0   & -     \\  
    ${\cal P}_{8}^{\dag}$  & 98.40  & -    & -    & 1.43 & 1.43 & 1.85   & -     \\
    ${\cal P}_{16}^{\dag}$ & 98.40  & -    & -    & 2.92 & 2.92 & 3.70   & -     \\
    ${\cal P}_{32}^{\dag}$ & 98.40  & -    & -    & 5.84 & 5.84 & 7.40   & -     \\
    ${\cal P}_{40}^{\dag}$ & 98.40  & -    & -    & 5.84 & 5.84 & 9.26   & -     \\    
    ${\cal P}_{48}^{\dag}$ & 98.40  & -    & -    & 8.75 & 8.75 & 11.11  & -     \\  
    \hline
    ${\cal P}_{8}$         & 98.40  & -    & -    & 1.43 & 1.43 & 1.81   & -     \\
    ${\cal P}_{16}$        & 98.40  & -    & -    & 2.92 & 2.92 & 3.70   & -     \\
    ${\cal P}_{32}$        & 98.40  & -    & -    & 5.84 & 5.84 & 7.40   & -     \\
    ${\cal P}_{40}$        & 98.40  & -    & -    & 7.24 & 7.24 & 9.16   & -     \\    
    ${\cal P}_{48}$        & 98.40  & -    & -    & 8.75 & 8.75 & 11.08  & -     \\
    \hline
    ${\cal U}_{08}$        & 108.60 & 3.79 & 2.25 & 1.47 & 1.47  & 1.86  & 0.66  \\
    ${\cal U}_{16}$        & 108.62 & 3.79 & 2.23 & 2.99 & 2.99  & 3.79  & 1.34  \\
    ${\cal U}_{32}$        & 108.65 & 3.79 & 2.24 & 5.75 & 5.75  & 7.28  & 2.58  \\
    ${\cal U}_{40}$        & 108.67 & 3.79 & 2.24 & 7.20 & 7.20  & 9.12  & 3.23  \\  
    ${\cal U}_{48}$        & 108.70 & 3.80 & 2.24 & 8.58 & 8.58  & 10.84 & 3.83  \\       
    \hline
    ${\cal S}_{8}$         & 108.61 & 3.79 & 2.24 & 1.47 & 1.89  & 1.86  & 0.66  \\
    ${\cal S}_{16}$        & 108.64 & 3.79 & 2.23 & 2.92 & 3.79  & 3.69  & 1.30  \\
    ${\cal S}_{32}$        & 108.66 & 3.79 & 2.24 & 5.49 & 7.32  & 6.91  & 2.45  \\  
    ${\cal S}_{40}$        & 108.69 & 3.79 & 2.24 & 7.29 & 9.69  & 9.24  & 3.27  \\  
    ${\cal S}_{48}$        & 108.72 & 3.80 & 2.24 & 8.73 & 11.48 & 11.03 & 3.90  \\      
    \hline
    ${\cal M}_{0}$         & 109.12 & 3.79 & 2.25 & 0.00 & 0.00  & 0     &  0    \\  
    \hline  
    ${\cal M}_{\eta}$      & 39.78  & 1.07 & 0.63 & 2.54 & 2.54  & 9.11  & 4.02  \\
    \hline
   \enddata     
   \label{tab:models} 
\end{deluxetable}


\subsection{The Echelle Diagrams for asteroseismic diagnostics\label{ssec:ED}}

One of the most frequently used techniques for seismic diagnostics of solar-like stars is the
representation of the oscillation frequencies in the so-called echelle diagram (hereafter, ED).
With such diagrams, one is able to identify the degree of modes and extract some information from
the asymptotic properties mentioned above.

Echelle diagrams are constructed by cutting the oscillation spectrum into frequency slices of size
$\alsep$, and stacking them up. Then the ED consists in depicting the oscillation frequencies as a
function of the same frequencies modulo $\alsep$. In a first-order asymptotic regime, EDs should be
composed of one vertical ridge per $\ell$, if modes are $m$-degenerate. When rotation is
considered, each vertical ridge corresponds to a given ($\ell,m$) mode.

\section{The seismic modeling \label{sec:modeling} }

The present work has been undertaken with the use of two types of models: 1D-2D polytropes and 1D
stellar models, the purpose and description of which are described in the following.

\subsection{Polytropes \label{ssec:smodels_poly}}

Polytropic models are used in this work for the comparison between perturbative and
non-perturbative approaches for the oscillation computations (section~\ref{sec:comppoly}).
This is so because, to date, no realistic (properly deformed by rotation) models of solar-like stars
are yet available. 

A polytropic model of index $n=3$ with $M=1.3\,\msun$ and $R=1.276\,\rsun$ was considered as the
reference model. From that model, 1D and 2D polytropic models with rotational velocities ranging
from $8\,\kms$ to $48\,\kms$ were built \citep[see][ for more details on 2D
polytropes]{Reese06}. The characteristics of all these models are listed in Table~\ref{tab:models}.

\begin{figure*}[!ht]  
  \centering
    \includegraphics[scale=0.43]{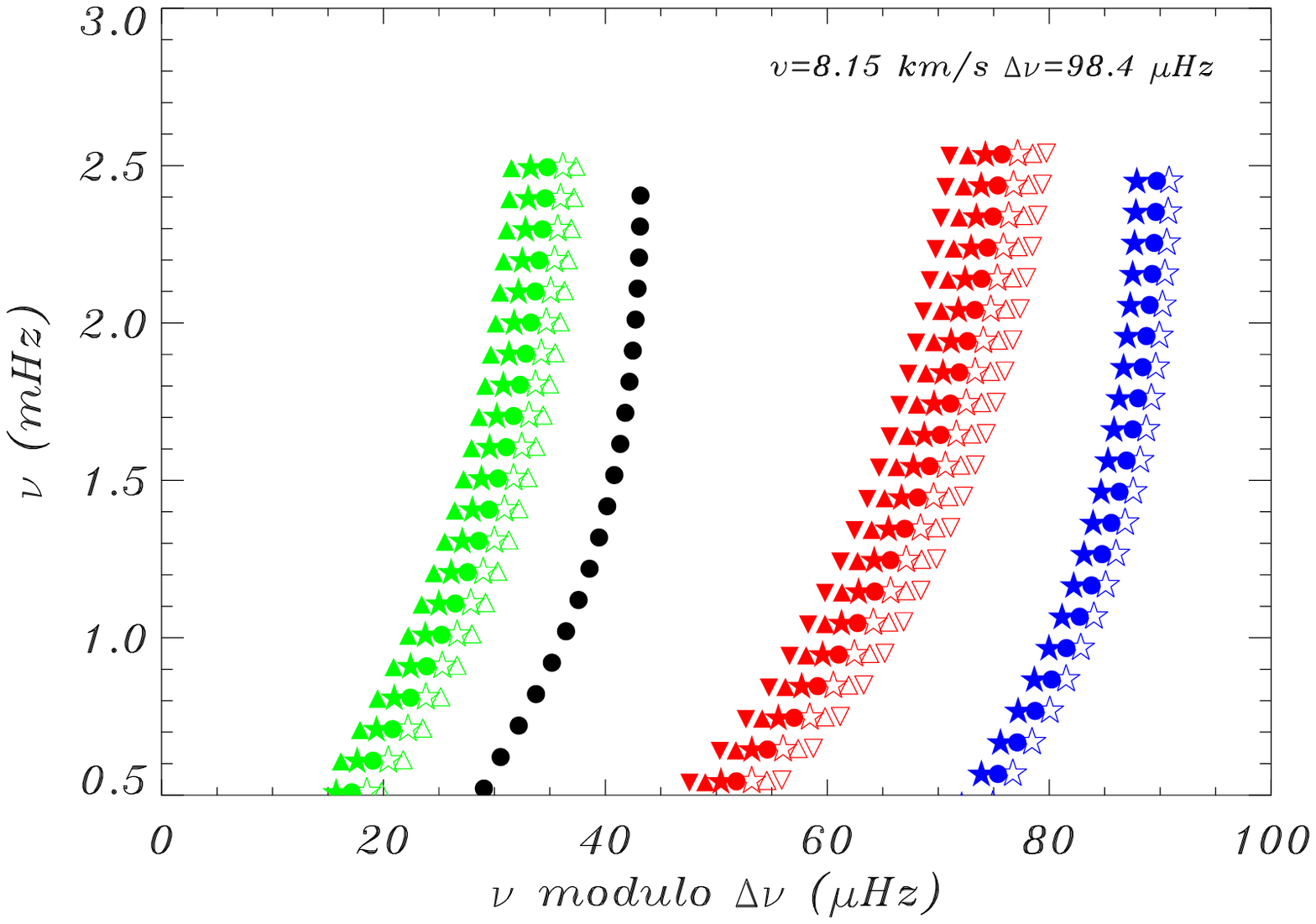} 
    \includegraphics[scale=0.43]{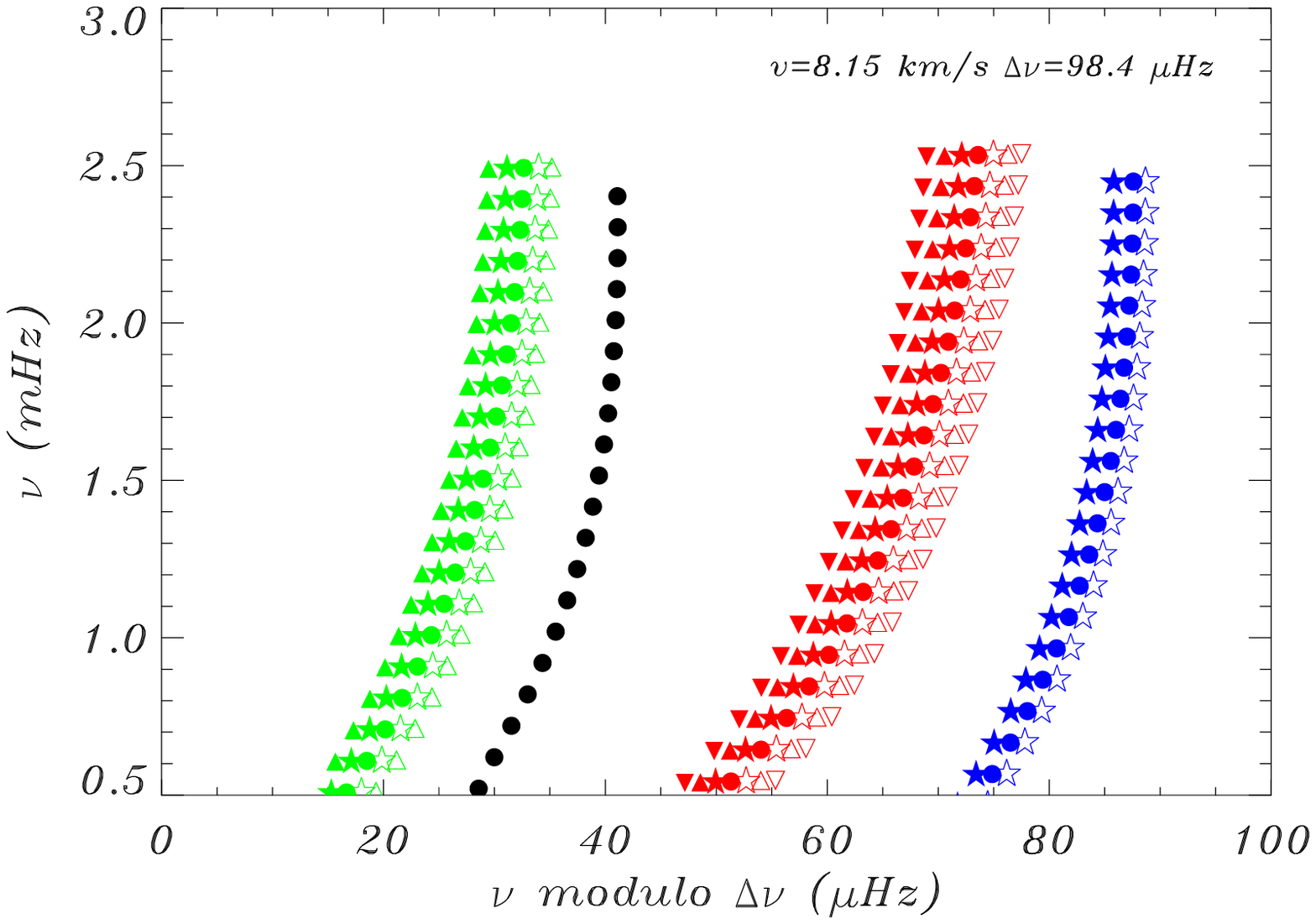}
    \includegraphics[scale=0.43]{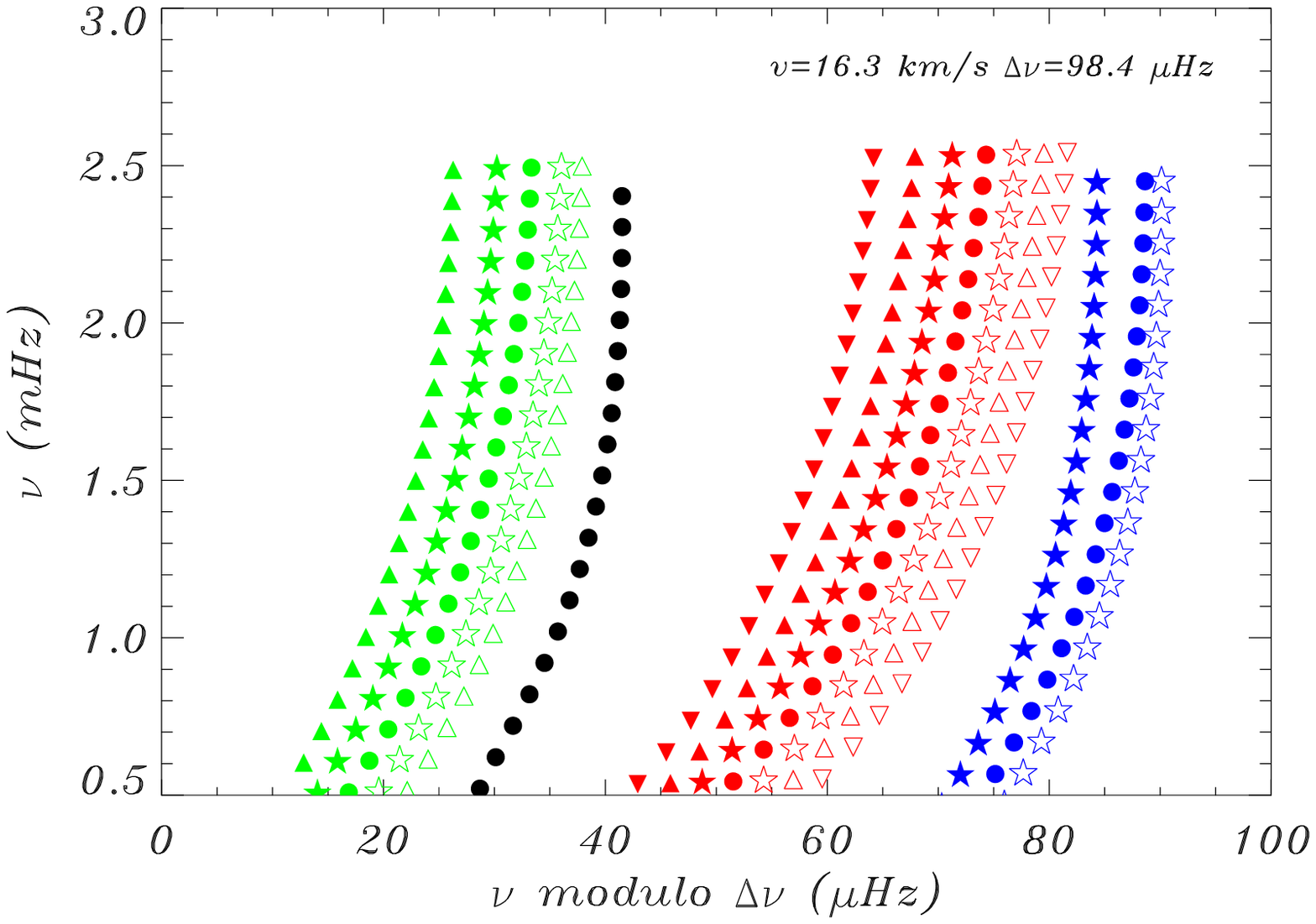} 
    \includegraphics[scale=0.43]{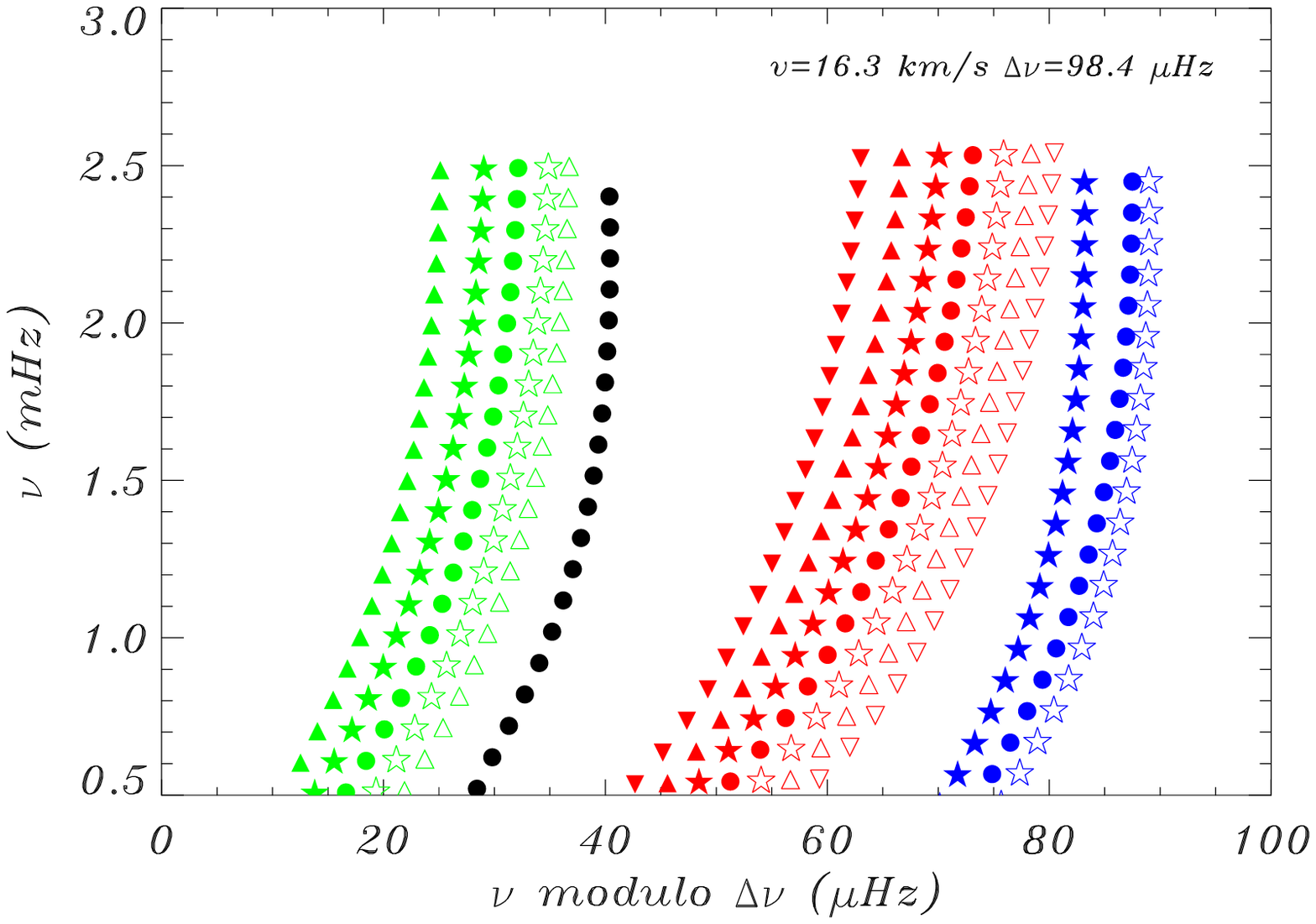}
    \includegraphics[scale=0.43]{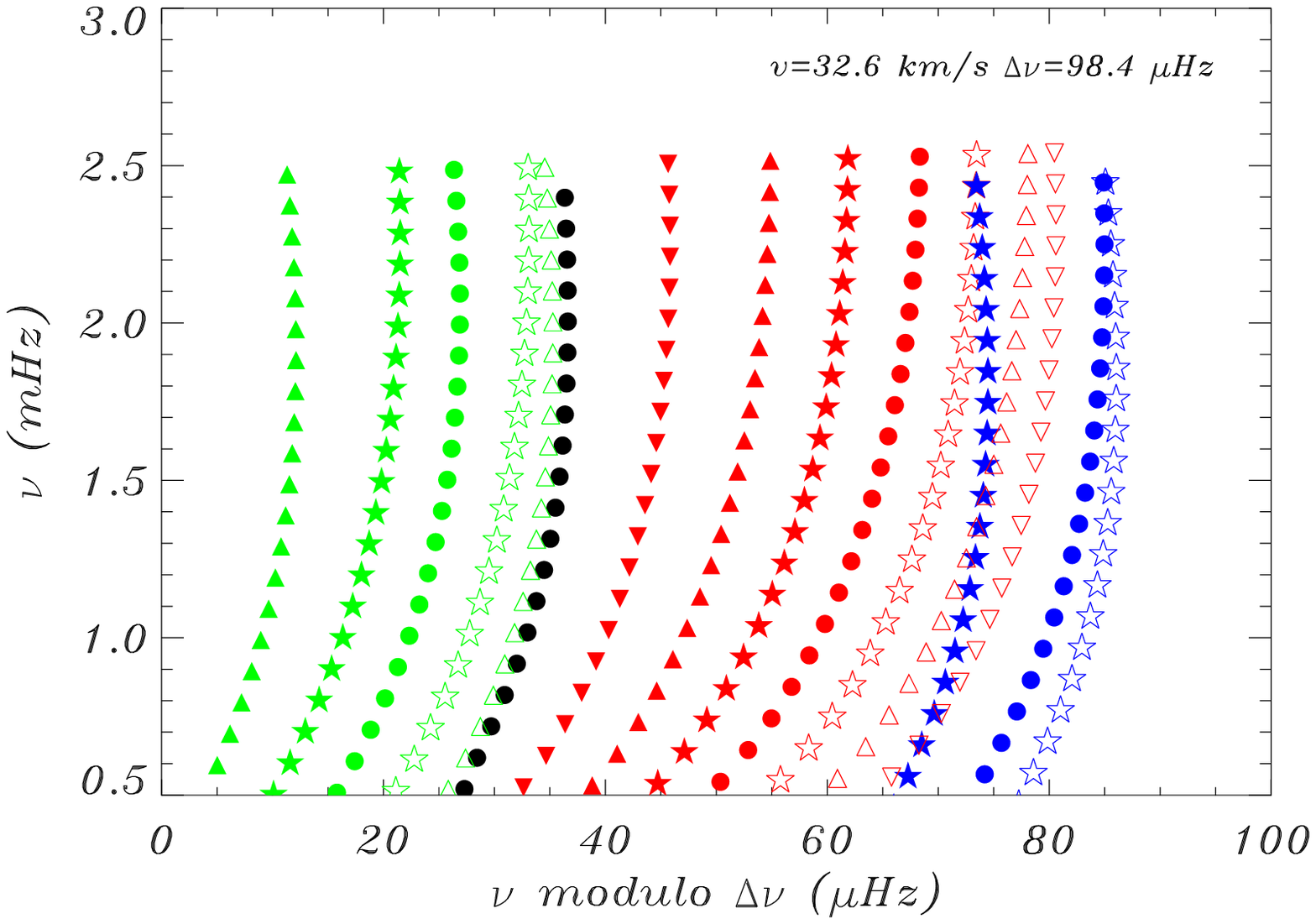} 
    \includegraphics[scale=0.43]{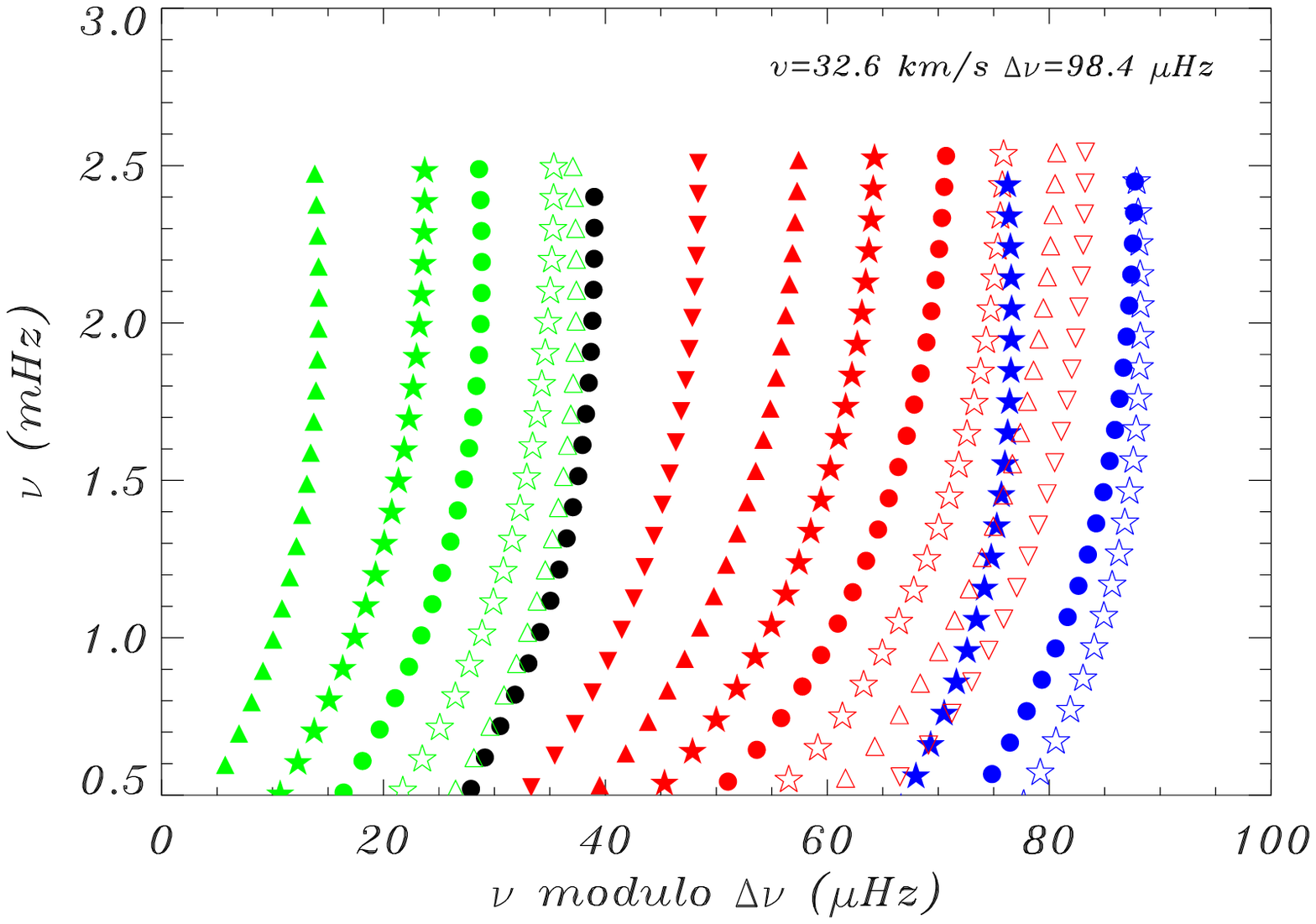}
    \includegraphics[scale=0.43]{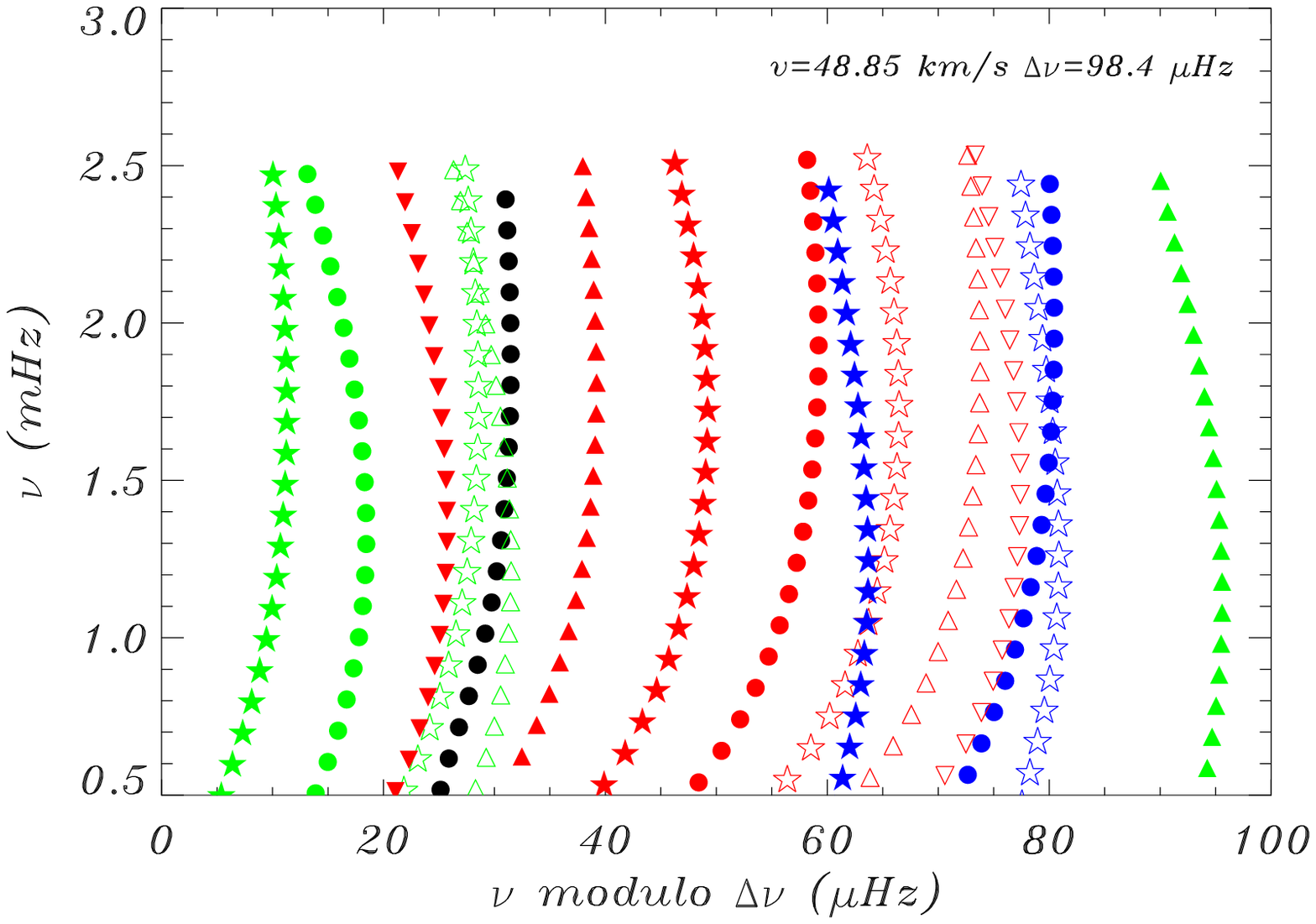} 
    \includegraphics[scale=0.43]{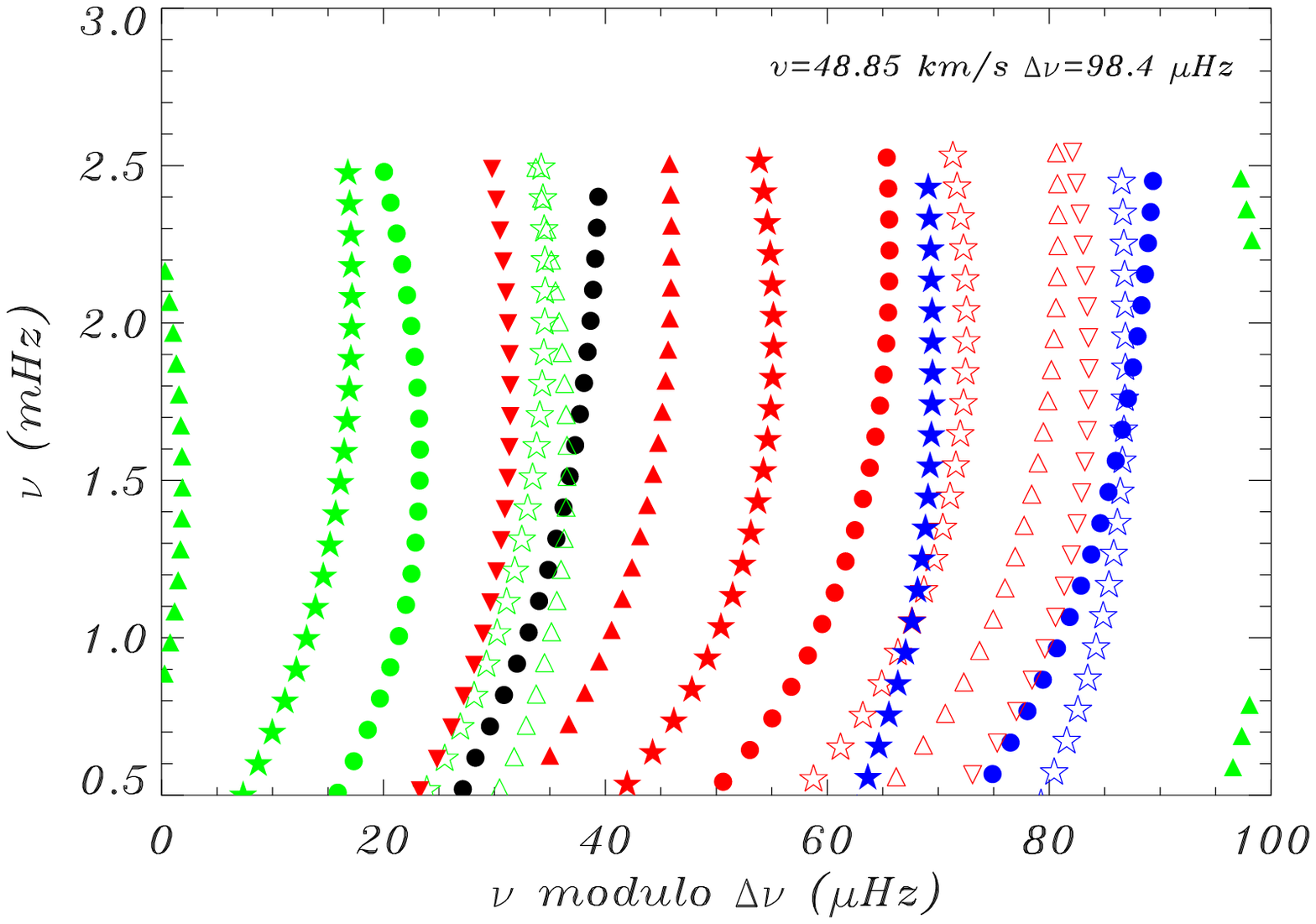}
    \caption{Perturbative (right panels) and non-perturbative (left panels) echelle diagrams
             computed for a polytropic model with four different rotational velocities (increase
             from top to bottom). Black and green symbols represent $\ell=0$ and $\ell=2$ modes,
             respectively. Blue and red symbols represent $\ell=1$ and $\ell=3$ modes,
             respectively. 
             Filled circles represent both $\ell=0$ and $m=0$ modes. For the rest of modes, filled
             and empty symbols represent the $-m$ and $+m$ frequencies, respectively. Stars,
             triangles, and inverted triangles represent modes with $|m|=1$, 2, and 3,
             respectively.}
\label{fig:edpolcomp}
\end{figure*}
\begin{figure*}[!ht]  
  \centering
    \includegraphics[scale=0.42]{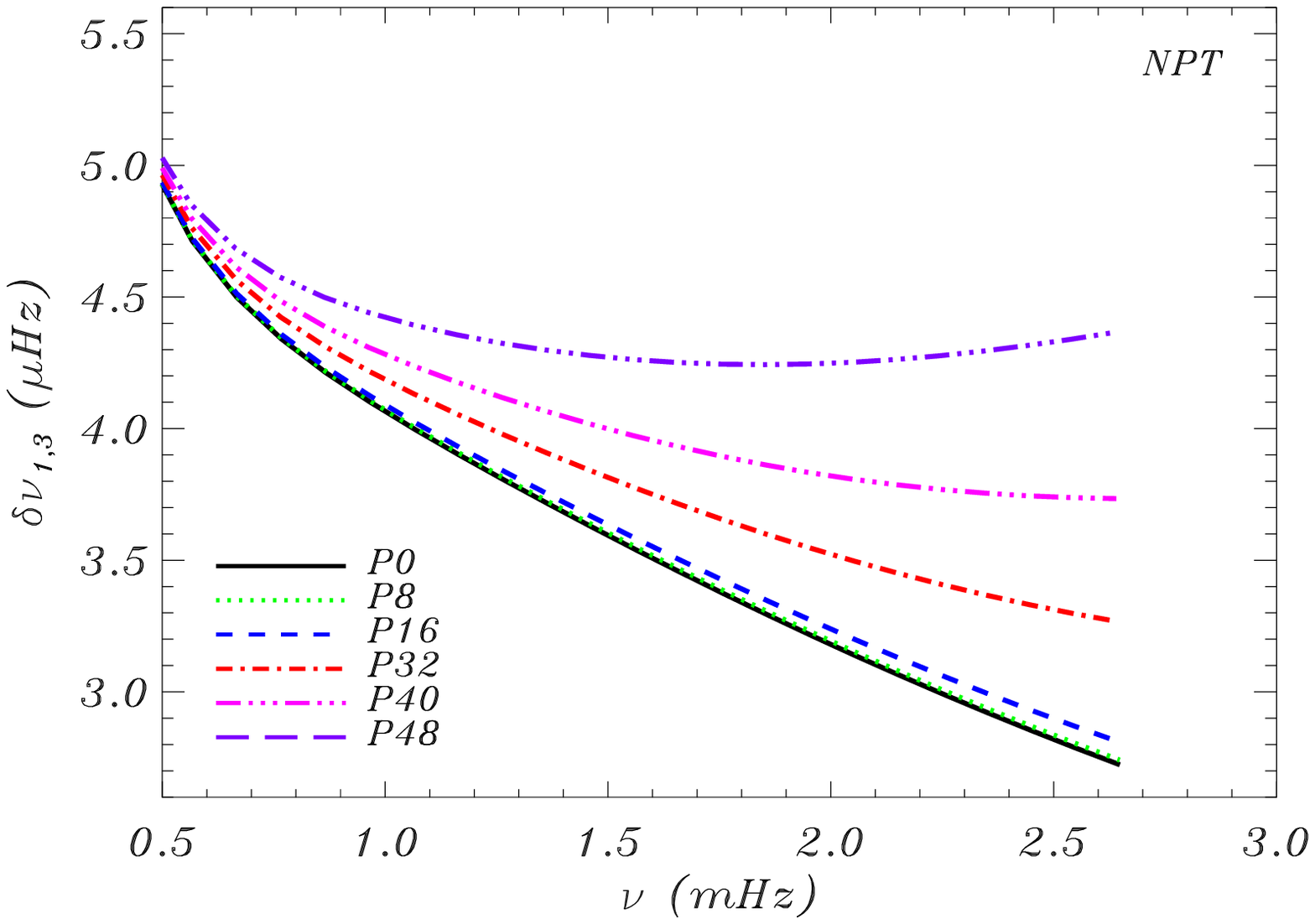}     
    \includegraphics[scale=0.42]{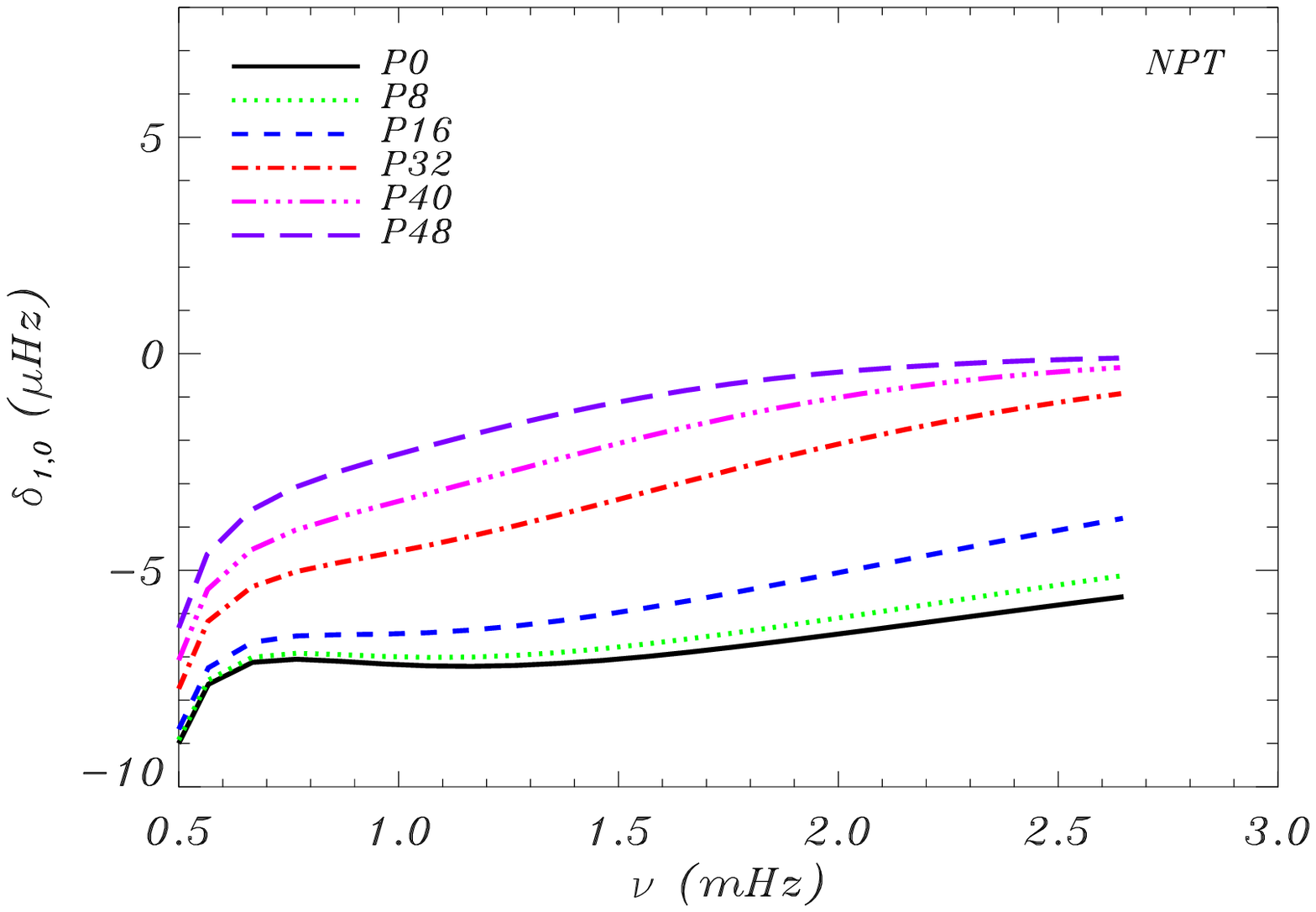}    
    \includegraphics[scale=0.42]{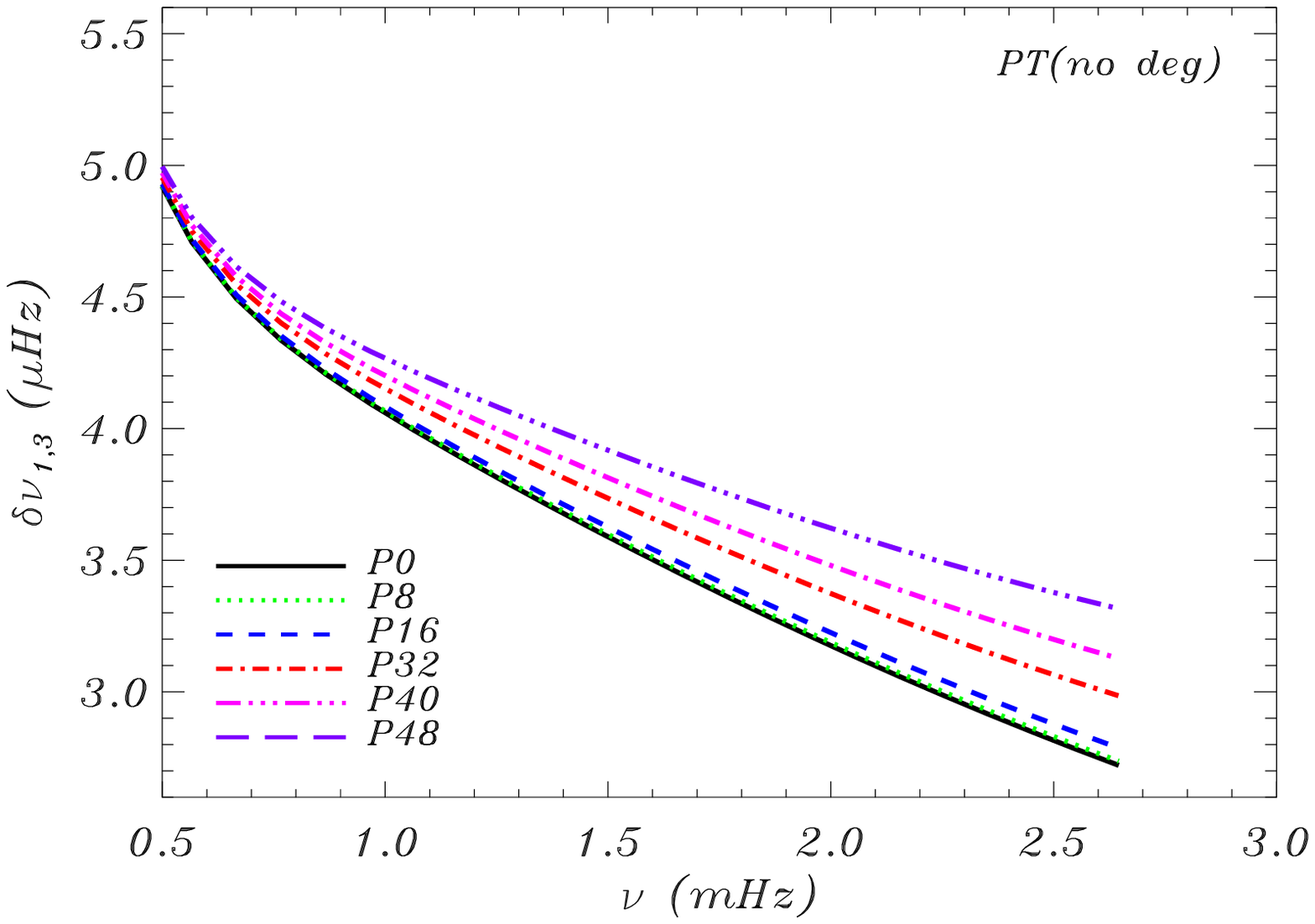}  
    \includegraphics[scale=0.42]{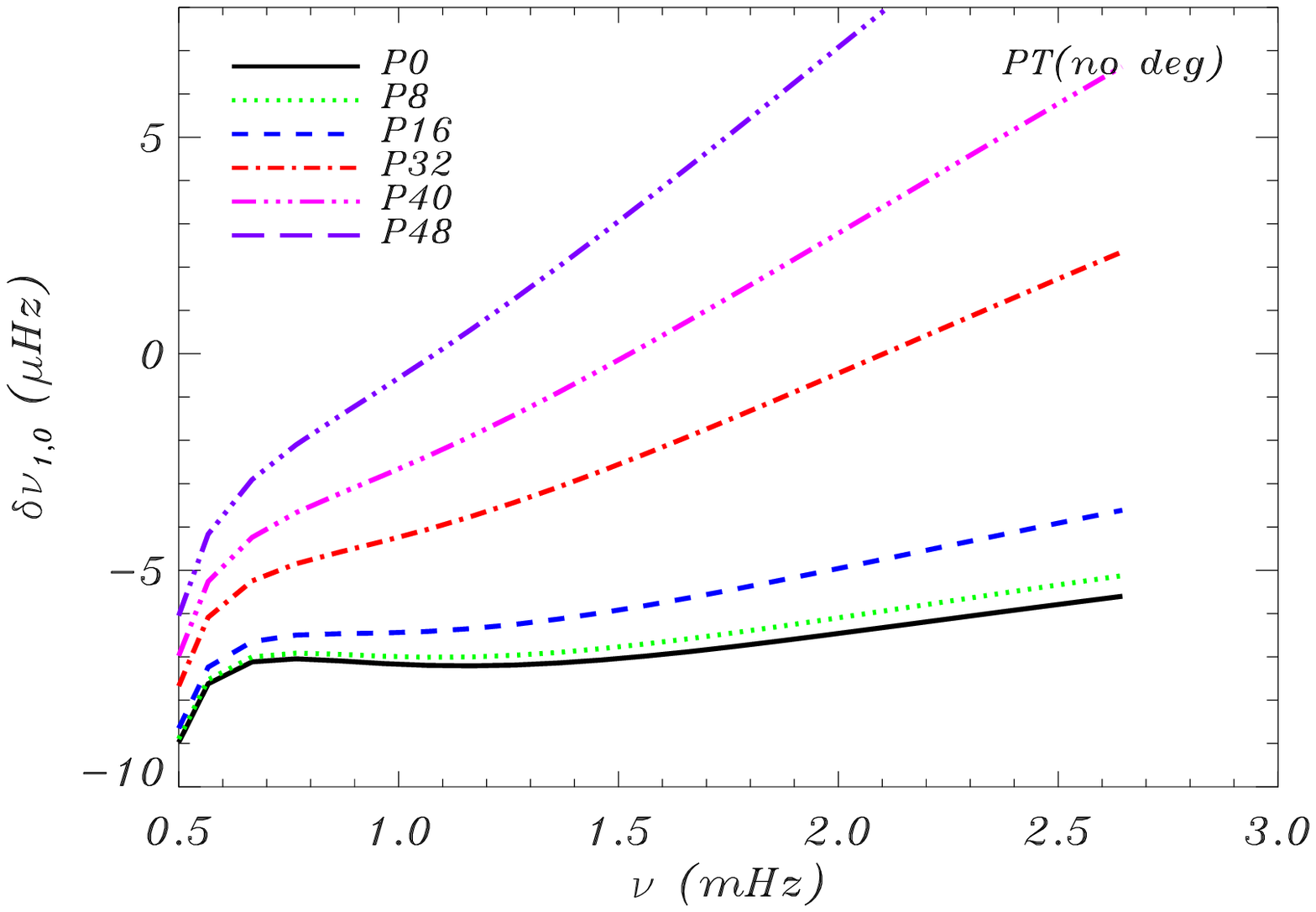}     
    \includegraphics[scale=0.42]{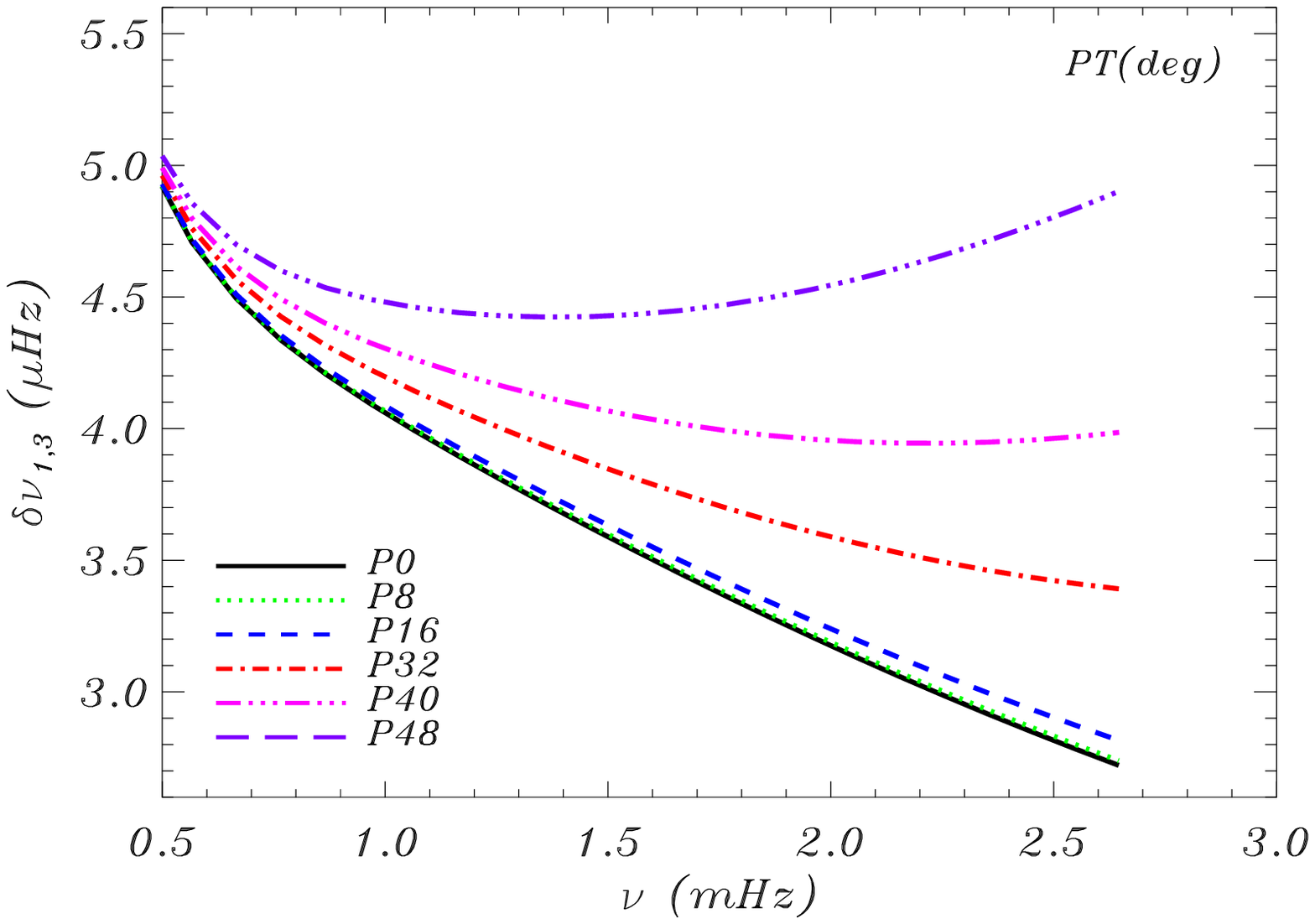}
    \includegraphics[scale=0.42]{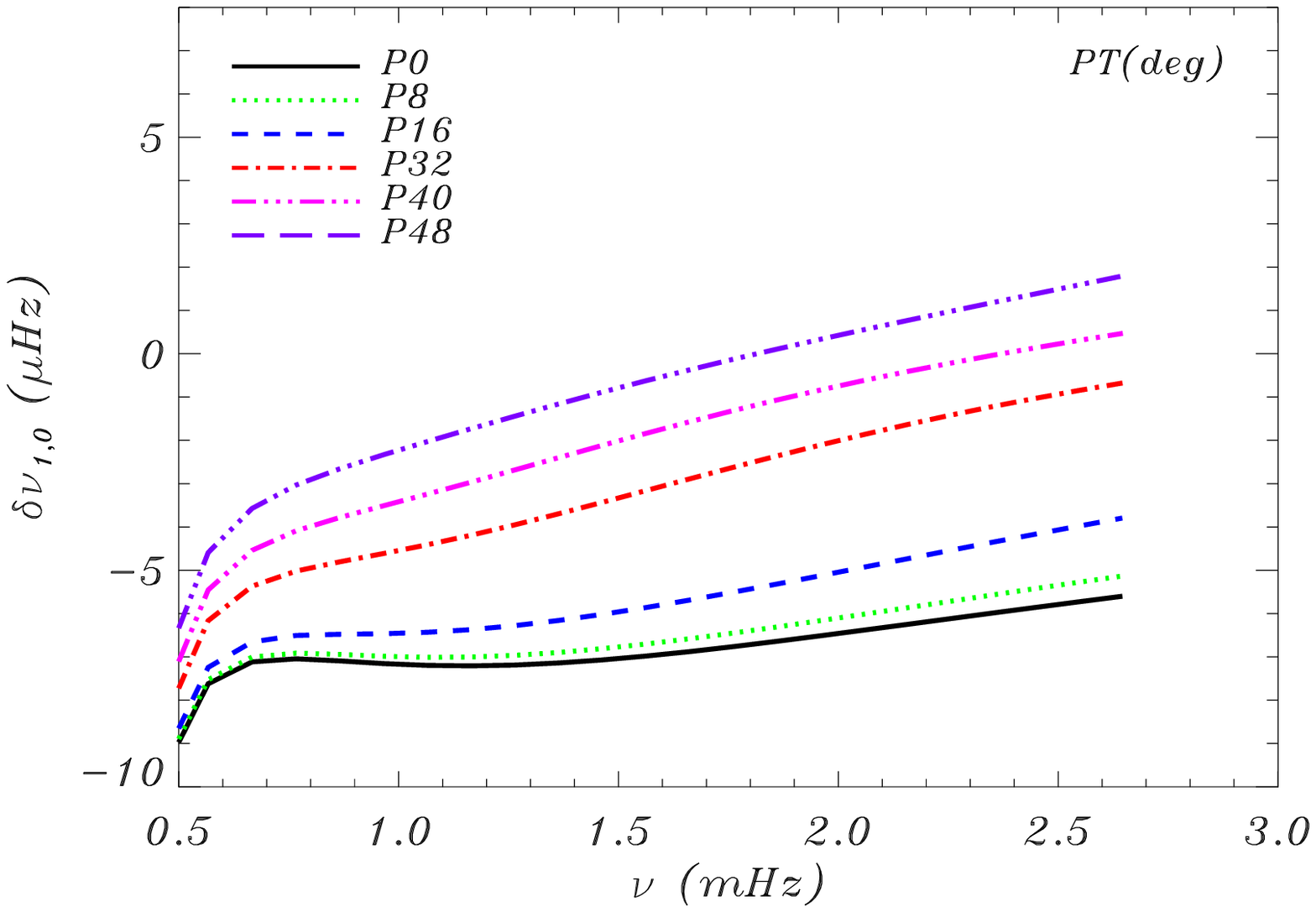} 
    \caption{Small spacings $\dnu_{0,2}$ (left panels) and $\dnu_{0,1}$ (right panels)
             as a function of the oscillation frequency. Panels on top correspond to
             NP oscillations. Middle panels show spacings calculated with a perturbative
             approach without including near-degeneracy effects. Bottom panels are similar to
             middle panels but near-degeneracy is included. Different line types 
             correspond to spacings computed for different rotational velocities.
             Colored figures are available in the on-line version of the paper.}
\label{fig:polspacings}
\end{figure*}

\subsection{Stellar models \label{ssec:rotmodels} }

Equilibrium models were computed with the \cesam\ code \citep{Morel97, MorelLeb08}.
Our stellar models include the spherically averaged contribution of the centrifugal acceleration 
through an effective gravitational acceleration in the hydrostatic equation according to 
\citet{KipWeig90}. Specifically, the  effective gravity is obtained as 
$g_{\mathrm{eff}}=g-{\cal A}_{c}(r)$, where $g$ is the local gravity, $r$ the radius, and
${\cal A}_{c}(r)=\frac{2}{3}\,r\,\Omega^2(r)$ the centrifugal acceleration \citep[similar models
have been previously used, for instance, by][]{Goupil00, Jagoda02, JCDThomp99, Sua06rotcel}.
Although the non-spherical components of the centrifugal acceleration are not considered in the
equilibrium models, they are included in the oscillation computations by means of a linear 
perturbation analysis described in \citet{DG92}, \citet{Soufi98},and \citet{Sua06rotcel}.

We consider two hypothesis about the internal rotation profile: uniform rotation (solid body
rotation), and shellular rotation (radial differential rotation)  with  local conservation of the
angular momentum during the evolution of the stellar model 
\citep[see][ for more details]{Sua06rotcel}. The characteristics of both uniformly- and
differentially-rotating models (${\cal U}_i$ and $\Sim$, respectively) are listed in
Table~\ref{tab:models}. We consider rotational velocities for the models spanning the range of
$\vsini$ observed in solar-like stars.

\subsection{Oscillations computation \label{ssec:oscill} }

Oscillations were computed using the non-perturbative and perturbative approaches. The former
consists in a numerical approach based on expansions of the equilibrium and oscillation variables on
spherical harmonics for the angular dependence, and on Chebyschev polynomials for the radial
dependence. Such computations were done for the 2D polytropic models described in
Section~\ref{ssec:smodels_poly}.

For the perturbative approach, the adiabatic oscillation code \filou\ was used
\citep{SuaThesis,Sua08filou}. This code corrects the oscillations frequencies up to the second-order
effects of rotation. These include near degeneracy effects, which occur when two or
more frequencies are close to each other. More specifically, in order to remain in the hypothesis
of the perturbative approach, we assume the loose condition that
 near degeneracy effects must be corrected for modes with frequencies satisfying:
\eqn{|\nu_{nlm}-\nu_{n'l'm})|\lesssim\frac{1}{2}|\alsep|.\label{eq:neardeg}}
In addition, the frequency computation  takes into account the presence of
a radial differential rotation profile of the form
\eqn{\Omega(r)={\bar \Omega}\,\Big(1+\eta_0(r)\Big)\label{eq:defOmega}}
where ${\bar \Omega}$ represents the angular rotational velocity at 
the surface and $\eta_0(r)$ a radial function representing \emph{shellular} rotation.
This profile is only considered in the radiative zone, in which the rotation rate decreases
approximately with a power of $r$, whereas instantaneous transport of the angular momentum is
assumed in the convective core, implying thus a uniform rotation profile in that part of the star
\citep[an illustration of such a profile can be found in ][]{Sua09apj} .

\section{Validity of the perturbative approach for oscillation computations\label{sec:comppoly}}
The perturbative approach is considered valid when the parameters
\eqna{\epsilon &=& \Omega/(G\,M/R^3)^{1/2}\\
          \mu  &=& \Omega/\nu_{n,\ell}\label{eq:defeps_mu}}
are small,that is, when (1) the stellar structure is not significantly
deformed by the centrifugal force, and (2) oscillation frequencies are much larger than the
angular rotation rate, respectively. More specifically, the stellar deformation, which scales
as $\epsilon^{1/2}$, needs to be small in comparison to the mode wavelength, which scales
as $1/\nu_{n,\ell}$. This implies that the use of a perturbative approach may fail when increasing
the rotational velocity of the star, and this failure is expected to come first for high
radial-order frequency modes, i.e. high frequency modes \citep{Lignieres06,Reese06}.

It is thus relevant to investigate the validity of the perturbative approach for the oscillation
computations, as well as to identify the possible effects that it may provoke on the echelle
diagrams. To do so, it is necessary to compare EDs constructed from synthetic oscillations computed
with both approaches. In order to try to remain within the nominal limits of validity of
perturbation theory, models were built with small values of $\epsilon$ and $\mu$, around $10^{-2}$
and smaller
(see Table~\ref{tab:models}).
\begin{figure}[!ht]
  \centering
    \includegraphics[scale=0.43]{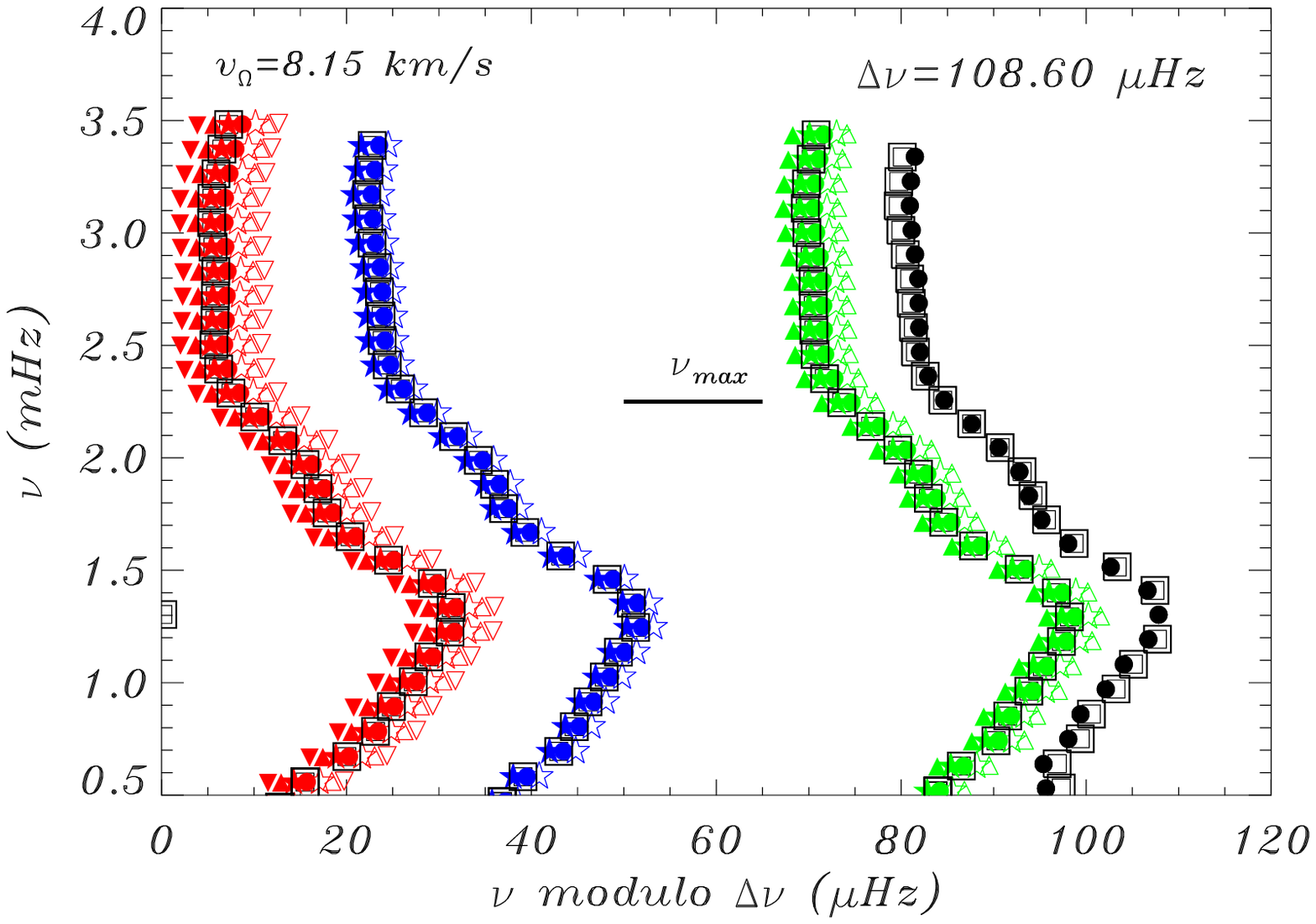}     
    \includegraphics[scale=0.43]{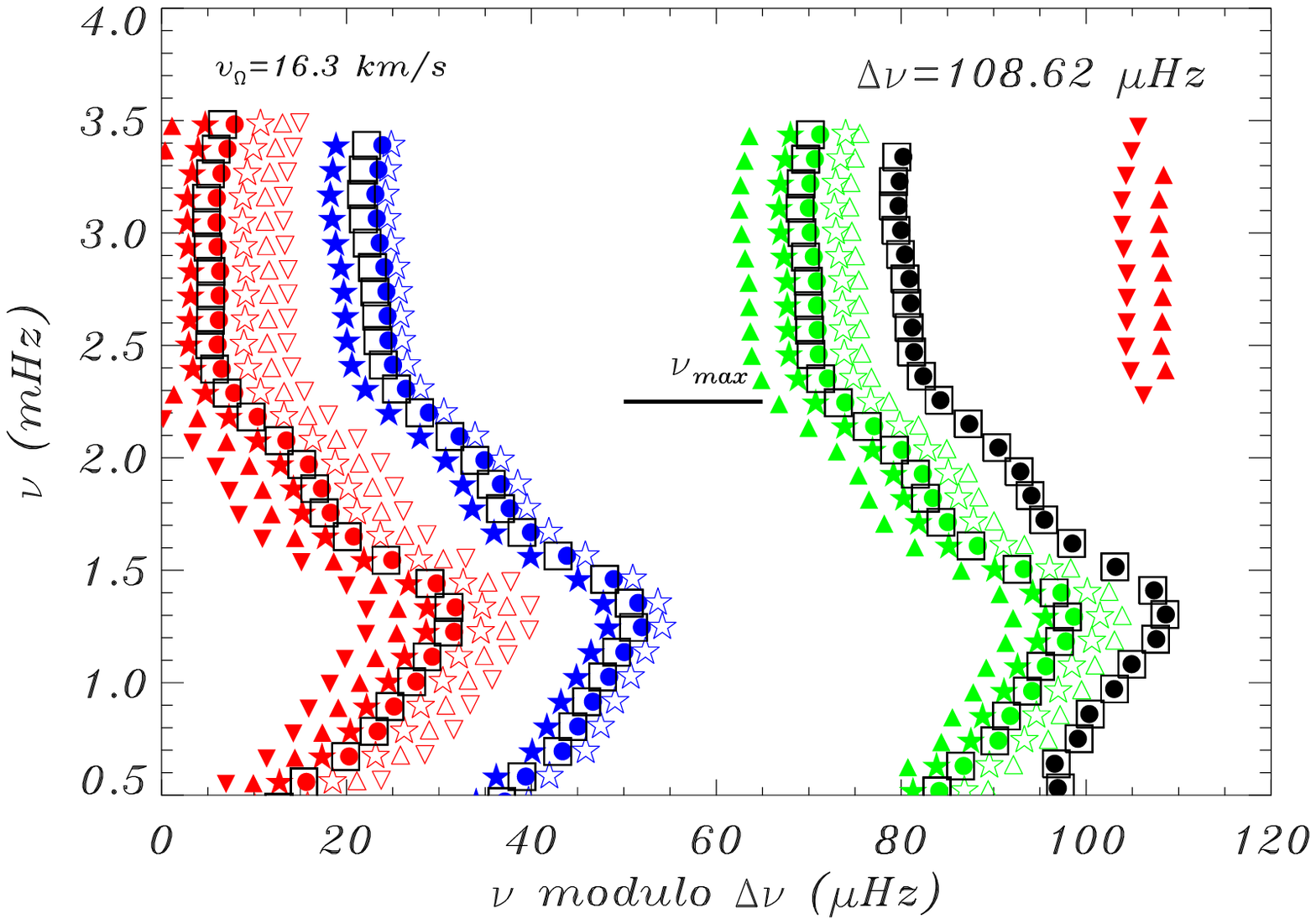}    
    \includegraphics[scale=0.43]{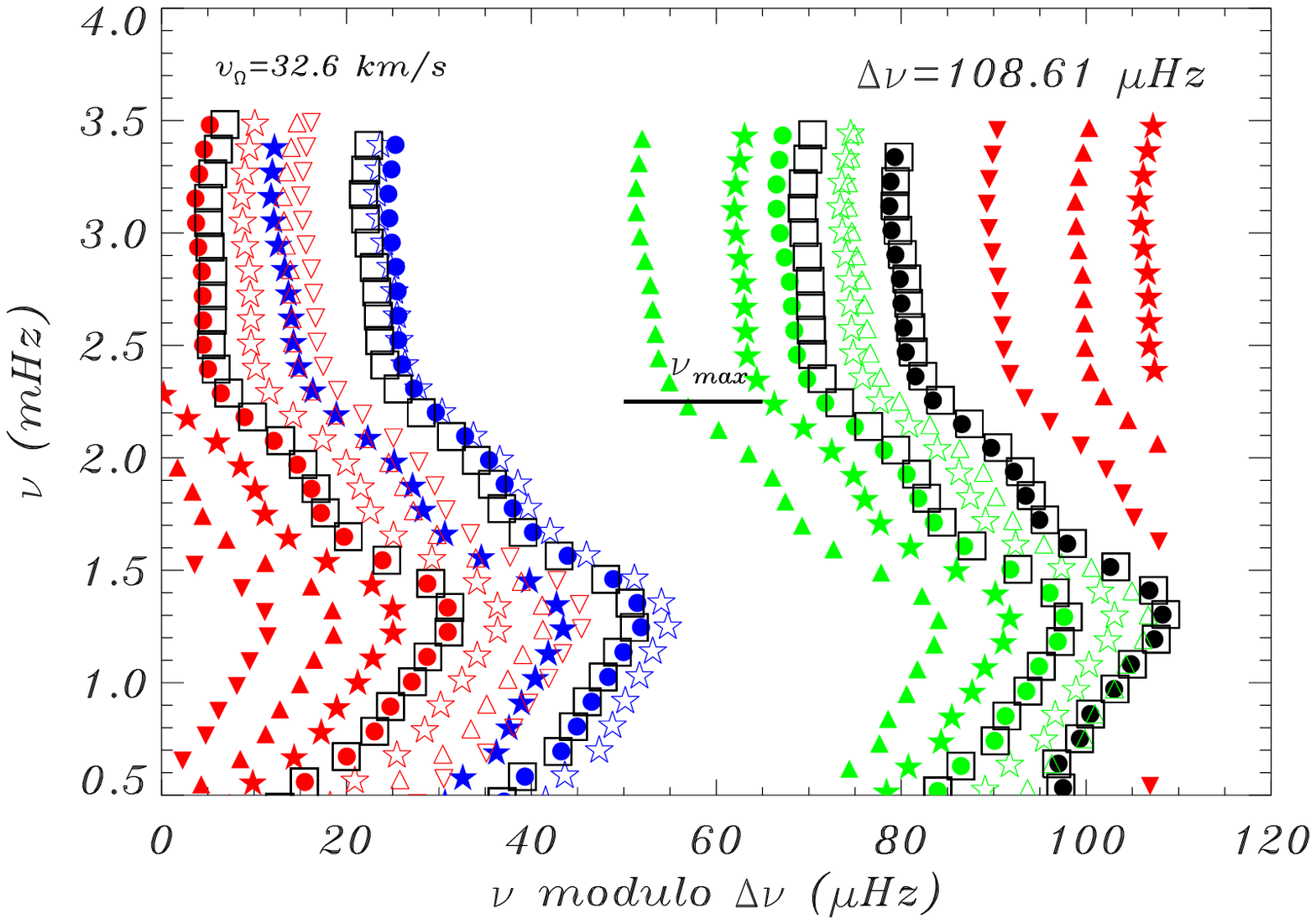} 
    \caption{Echelle diagrams for the uniformly-rotating models $\Uim$. Open squares correspond 
             to ridges of the non-rotating stellar model ${\cal M}_{0}$. The remaining symbols
             and colors have the same meaning as in Figure~\ref{fig:edpolcomp}. The small horizontal
             line shows the location of the $\numax$ frequency. Stars,
             triangles, and inverted triangles represent modes with $|m|=1$, 2, and 3,
             respectively.(For clarity, colors are used in the
             \emph{on-line} version of the paper).}
\label{fig:edrurd}
\end{figure}

For the sake of shortness and simplicity, P and NP will indicate the perturbative and
non-perturbative cases, respectively. For instance, echelle diagrams computed using the perturbative
approach will be denoted by PEDs (Perturbative Echelle Diagrams), and those computed with a
non-perturbative approach will be denoted by NPEDs (Non-Perturbative Echelle Diagrams). For both P
and NP models, uniform rotation is assumed.

The comparison between both types of diagrams including all the $m$ components of modes shows that
these are very similar for rotational velocities lower than $32\,\kms$ approximately (see
Figure~\ref{fig:edpolcomp}). For higher rotational velocities ($\epsilon\gtrsim0.09$), the P results
should be interpreted with caution.

Nevertheless only slight differences (for this scaling, i.e. a few $\muHz$) between PED and NPEDs
come out for $\ell=0$ and $\ell=1,m=0$ modes in the high-frequency region ($\nu_{n,\ell}>2\,\mHz$). 
This means that, at the scale of echelle diagrams, both P and NP can be considered as
quasi-identical. This is also the case when comparing the P and NP large frequency spacing.

For the small spacings, we find very similar results from both approaches, as can be deduced from
the comparison between the top and bottom panels in Figure~\ref{fig:polspacings}. For
$\dnu_{0,2}$, P spacings show the same behavior as a function of the frequency as their NP
counterparts. Indeed, the results are quasi-identical for models with rotational velocities up to
$40\,\kms$. Deviations of P with respect to NP spacings are more evident for $48\,\kms$, showing
differences of around $1\,\muHz$, which are visible for frequencies higher than $1.5\,\mHz$ and
which increase with the oscillation frequency. Similarly, the analysis of P and NP results for
$\dnu_{1,0}$ are similar, with increasing differences from $1.5\,\mHz$ to higher frequencies.

Furthermore, when near degeneracy effects are not considered, the resulting small spacings deviate
from the NP results even for the lowest rotational velocities considered here. This is shown in
Figure~\ref{fig:polspacings} (middle panels), where it can be seen that such deviations increase
with the frequency and the rotational velocity. This can be explained by the dependency of the
near-degeneracy correcting terms
upon, among other terms, the frequency, the rotation profile, and structure terms affected by the
centrifugal force \citep{Sua06rotcel}. This issue, together with a more detailed and quantitative
comparison, is currently in progress, and the results will be published in a forthcoming paper 
\citep[][ in prep.]{Ouazzani10prep}. 

Several important consequences can be extracted from these results: (1) It is correct to use the P
approach for studying the effect of rotation on EDs (up to $\epsilon=0.09$), (2) even for low
rotational velocities, second-order effects of rotation, including near degeneracy effects, must be
included in the oscillation computations, and (3) there are no artificial features in the PEDs which
could jeopardize the correct physical interpretation of the seismic diagnostics performed with them
(even at a small scale) as long as near degeneracy effects are properly taken into account.

\section{EDs of stellar models \label{sec:ED_sm}}

Following the conclusions on the validity of the perturbation methods given in the previous section,
we computed the oscillations of the $\Uim$ and $\Sim$ models (Table~\ref{tab:models}) including
near degeneracy effects. It can be shown that EDs of $\Sim$ and $\Uim$ models are very similar, so
for simplicity, discussion will be focused on $\Uim$  models. For brevity, ridges in the EDs are
specified with the notation ($\ell,m$).

Figure~\ref{fig:edrurd} shows the echelle diagrams calculated for the three $\Uim$ models of
Table~\ref{tab:models}. At first sight, one notices that second-order effects of rotation do not
change substantially the overall shape of the EDs, which remains dominated by the internal
stratification of the star and the presence of hydrogen and helium ionization zones. However,
significant variations (within the global ED structure) are found for increasing rotation rates
$\Omega$, and for different frequency domains. In fact, such variations should be considered in
terms of ($\epsilon,\mu$) values, rather than a function of the rotational velocity. It can
be shown, for instance, that for similar rotational velocities but different masses, EDs show
different results. Here, for clarity, the discussion is nevertheless held with respect 
to $v_{{\rm rot}}$.

For ${\cal U}_{08}$, the multiplets with different $\ell$ remain quite distant from each other. 
In particular the frequency gap between the $\ell=0$ and 2 multiplets is around $20\,\muHz$, and
between the $\ell=1,3$ multiplets around $10\,\muHz$. This roughly corresponds to the spacing
between $m=0$ ridges for those multiplets. For such a low rotational velocity, first-order effects
of rotation, which are proportional to $m\Omega$, dominate, thereby explaining the symmetry of the
multiplets. In this case, the rotational splitting is $1.47\,\muHz$ (see Table~\ref{tab:models}),
which represents approximately the 7\% and 14\% of the average distance between the 0-2 and 1-3
multiplets, respectively, and even more importantly, about 1\% of the ED scale ($120\,\muHz$). 

\begin{figure*}[!ht]
  \centering
    \includegraphics[scale=0.43]{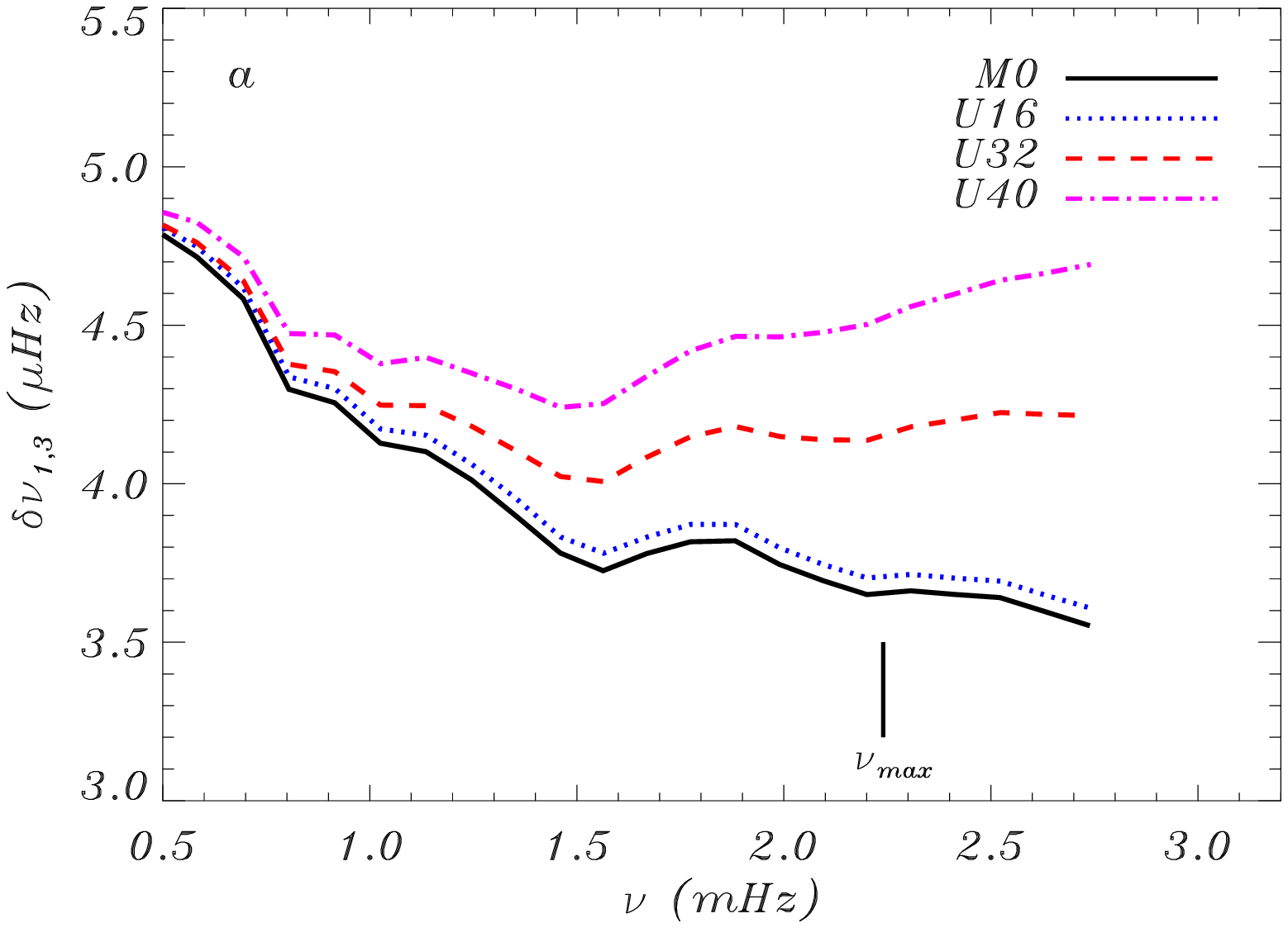}     
    \includegraphics[scale=0.43]{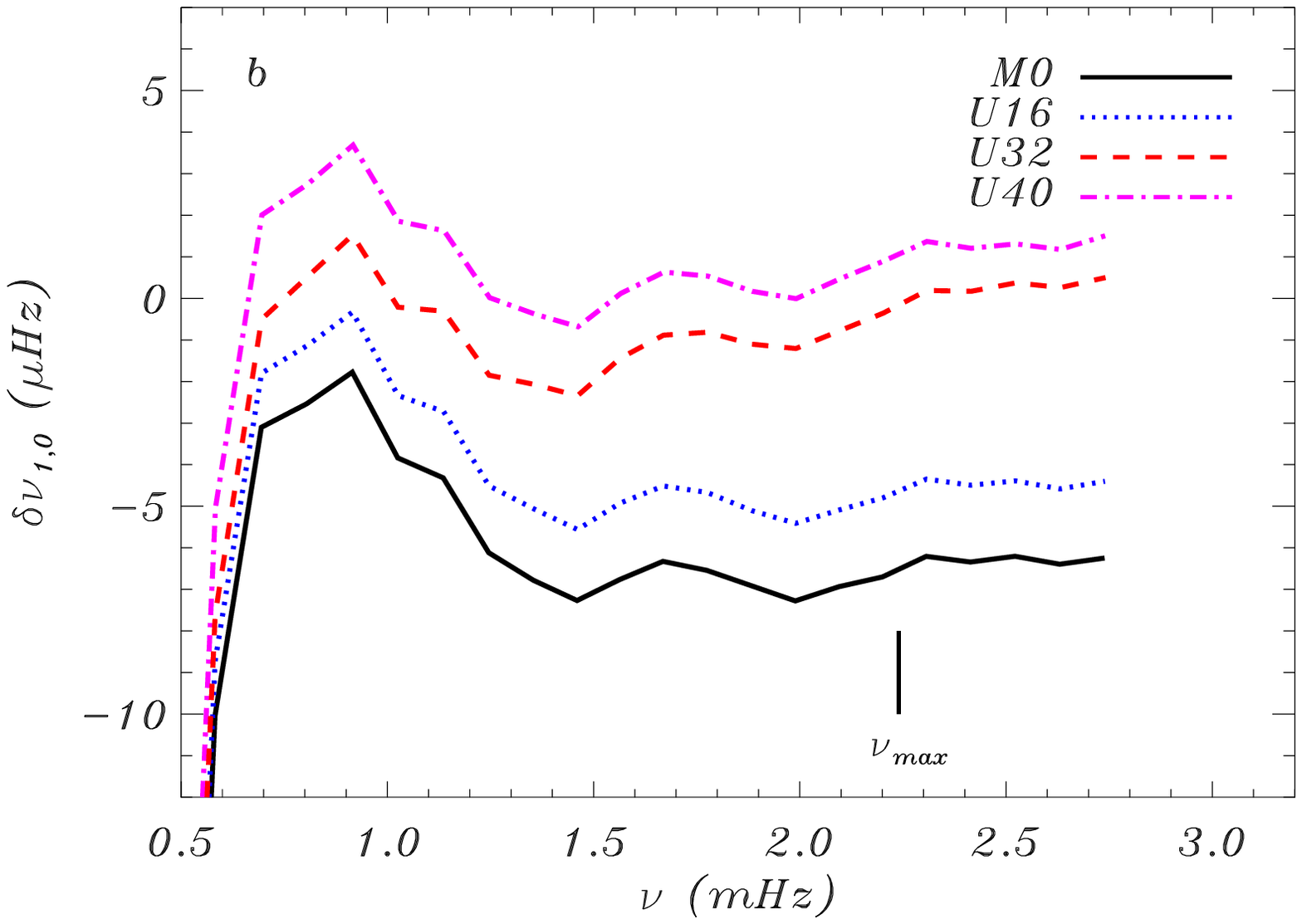}       
    \includegraphics[scale=0.43]{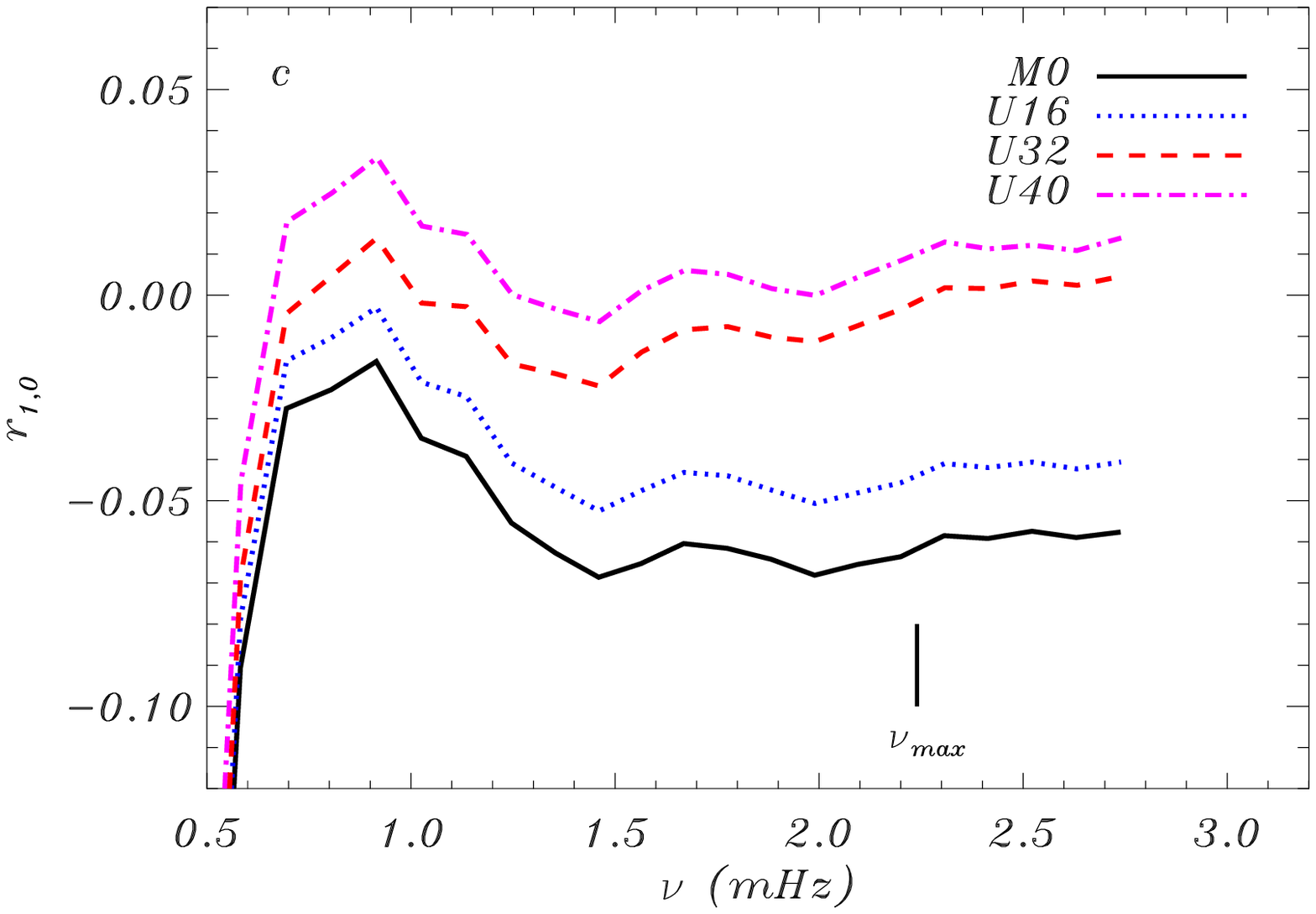}       
    \includegraphics[scale=0.43]{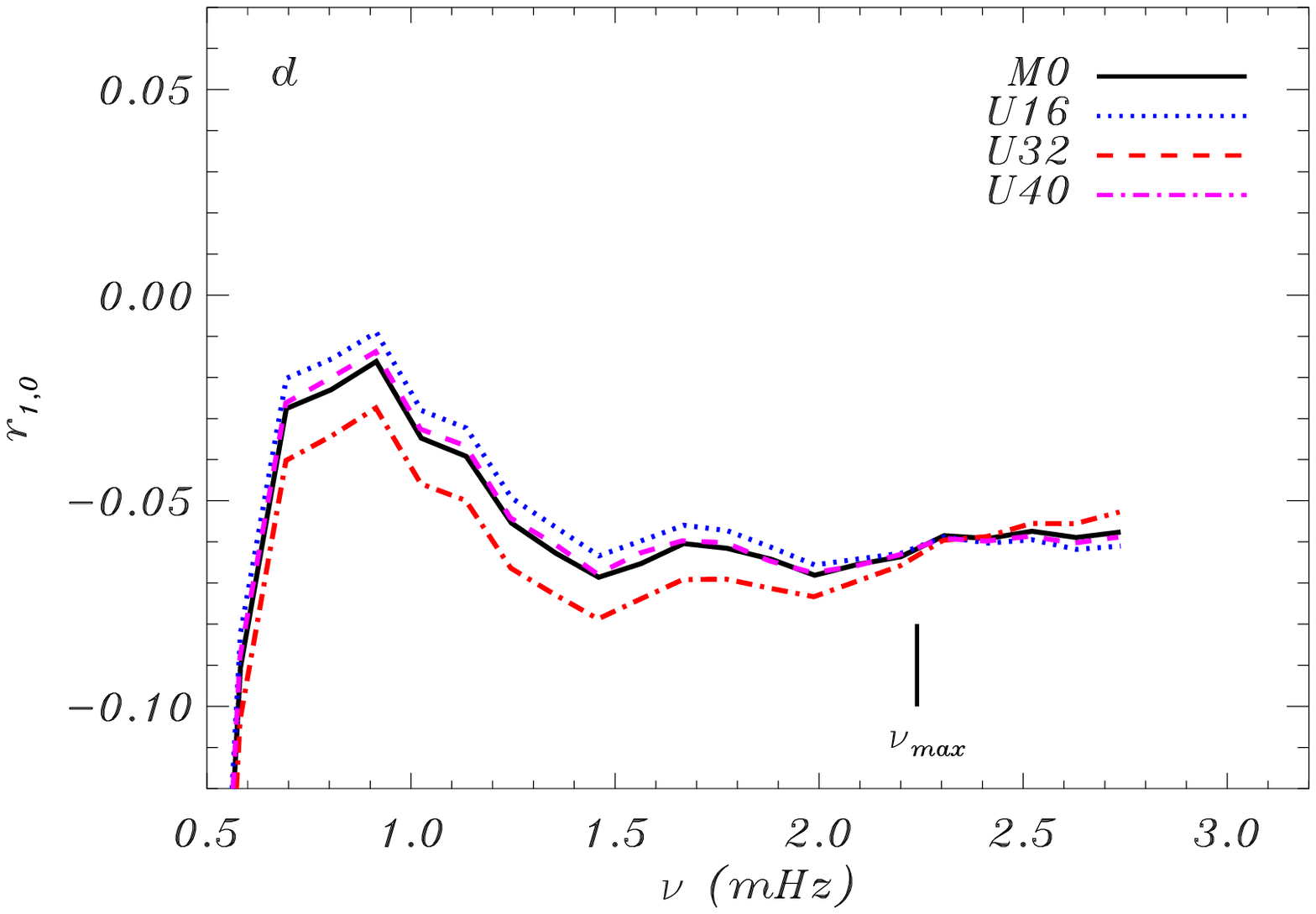}      
    \caption{Effects of rotation on different small frequency spacings represented as 
             function of the oscillation frequency. Panel (a), and (b) show the small
             frequency spacings $\dnu_{1,3}$ and  $\dnu_{1,0}$, respectively. Panel
             (c) and (d) show the scaled spacings $r_{1,0}$, the latter being 
             calculated without taking second-order effects of rotation into account.
             The small vertical line indicates the location of the $\numax$ frequency.}
\label{fig:spacings}
\end{figure*}

For ${\cal U}_{16}$, however, the symmetry of multiplets is broken for frequencies greater than
$1.5\mHz$ approximately, indicating that, from rotational velocities around $16\,\kms$, second-order
effects of rotation become important. Nevertheless, such effects are not strong enough to mix up
multiplets of different mode degrees, whose centroid modes are separated (in average) by about
$10 \,\muHz$(0-2 multiplets) and $20\,\muHz$ (1-3) multiplets. However, the rotational splitting for
this model is $3\,\muHz$ and the $m$ ridges are almost two times (in average) more spread in the ED
than those of model ${\cal U}_{08}$. Consequently, the identification of the mode degree is not so
obvious in this case: The (1,+1) ridges are closer to (3,-3) as the frequency increases (around
$4\,\muHz$ at $\nunl\backsimeq1.5\mHz$, and $2\,\muHz$ at $\nunl\backsimeq3\mHz$). Similarly, radial
modes ridges are close to (2,-2) ones (less than $5\,\muHz$). This is particularly relevant for the
highest frequency domain.

But the most extreme case considered here is the ${\cal U}_{32}$ model. With such a large
rotation rate, second-order effects including near degeneracy effects
(see Section~\ref{ssec:rotmodels}) become significant. From Figure~\ref{fig:edrurd}, one notices
that
there is a blend of ridges which might blur the typical ED features, in particular when few modes
are detected. The scenario described for ${\cal U}_{16}$ is somehow intensified now. The blending is
caused by ridges approaching, and even crossing each other in certain cases: The (1,-1) ridge
crosses (1,0) at $\nunl\backsimeq1.5\mHz$. Similarly, (2,-2) crosses the radial modes 
at $\nunl\backsimeq1.8\mHz$. Moreover, (3,$-m$) are very close to any of the (1,$m$)
ridges, e.g. (3,-1) with (1,+1), (3,-2) with (1,+1), and (3,-3) with both (1,+1) and (3,+2).
In such a  case, unless one has good independent reasons to assume that only m=0 modes are detected,
it would be extremely difficult to attribute a ($\ell,m$) value to some of the modes.

\section{Discussion \& conclusions  \label{sec:discussion}}

We showed that that even for mild rotation, second-order effects of rotation start to have
significant impact on the oscillation frequencies in the range of their expected maximal detection,
that is around $\nu_{max}$, i.e. the frequency at which the observed power spectrum reaches its
maximum. Here, we use the empirical relation calibrated to solar values \citep{Bedding04} can be
written as
\eqn{\numax=\numaxsun\frac{\nuac}{\nuacsun},\label{eq:nu_max}}
where $\nuac$ is the acoustic cut-off frequency defined by
\eqn{\nuac = c/\Hp\label{eq:defnuac}}
where $\Hp$ is is the pressure scaleheight. We assumed the value for the
solar calibration $\nuacsun/\numaxsun\sim1.70$  \citep{BalGough90}. Indeed, rotation effects on EDs
(see Figure~\ref{fig:edrurd}) are of the order of the $\muHz$ for rotational velocities higher that
$8\,\kms$  ($\epsilon\sim0.018, \mumax\sim0.6$) for frequencies around $\numax$ (see
Table~\ref{tab:models}) and higher.

\subsection{Mode identification \label{ssec:modeid}}

From the above results, one question naturally arises: How can we distinguish between different $m$
ridges when they are close or even crossing each other? This question has
no easy answer. There are many physical variables that must be considered when analyzing mode
excitation and detection: geometry, visibility, intrinsic amplitudes, excitation mechanisms, etc.
It is not the purpose of this work to go into great detail on these aspects, however, analysis of
Fig~\ref{fig:edrurd} can provide some  hints which could help with mode identification. For
instance, notice that $-m$ ridges spread more drastically than the $+m$ ones, this effect being
more significant as the rotation velocity increases. This would thus help to identify the $m\neq0$
components of the oscillation spectra, in particular in the case for which this differential
behaviour is stronger, i.e. for the $\ell=|m|$ modes. In fact, based on visibility and geometrical
considerations, those modes are more visible for increasing inclination angles, the limit case 
being $i\rightarrow 90\deg$, that is when the star is being observed equator-on. Furthermore, it has
been found empirically that observed amplitudes for p modes in main sequence stars are correlated
with increasing $i$ \citep{Sua02aa}. This result was then supported by theoretical simulations using
non-perturbative theory for the oscillation computations \citep{Lignieres06} which predict that the
mode amplitude tends to concentrate near the equator \citep[firstly predicted by][]{Clement98}. It
can thus be concluded that the larger
the rotational velocity of the star is, the more chances the observed ridges in EDs
correspond to $\ell=0$ and/or $\ell=|m|$ modes. 
\begin{figure*}[!ht]  
  \centering
    \includegraphics[scale=0.43]{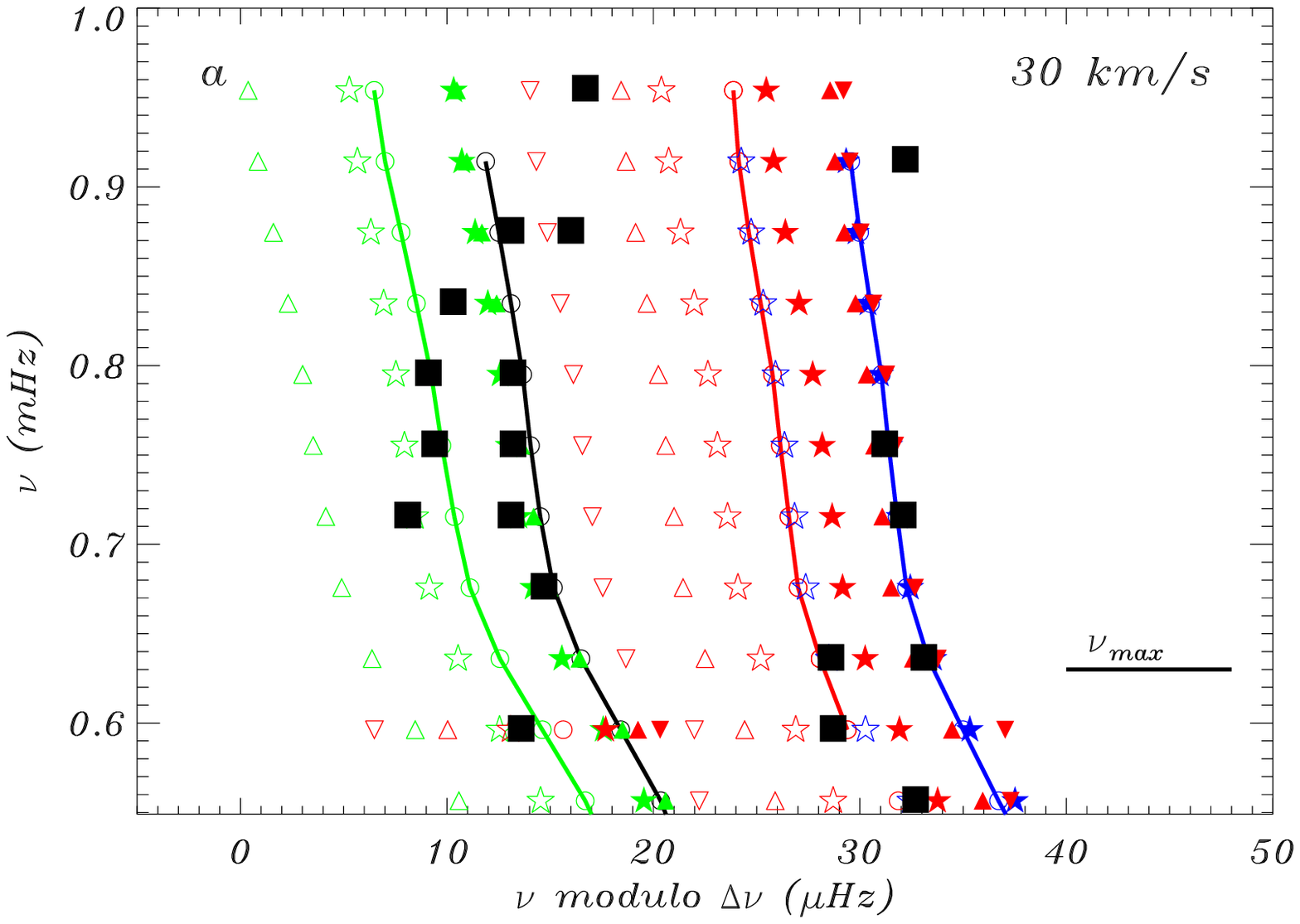}
    \includegraphics[scale=0.43]{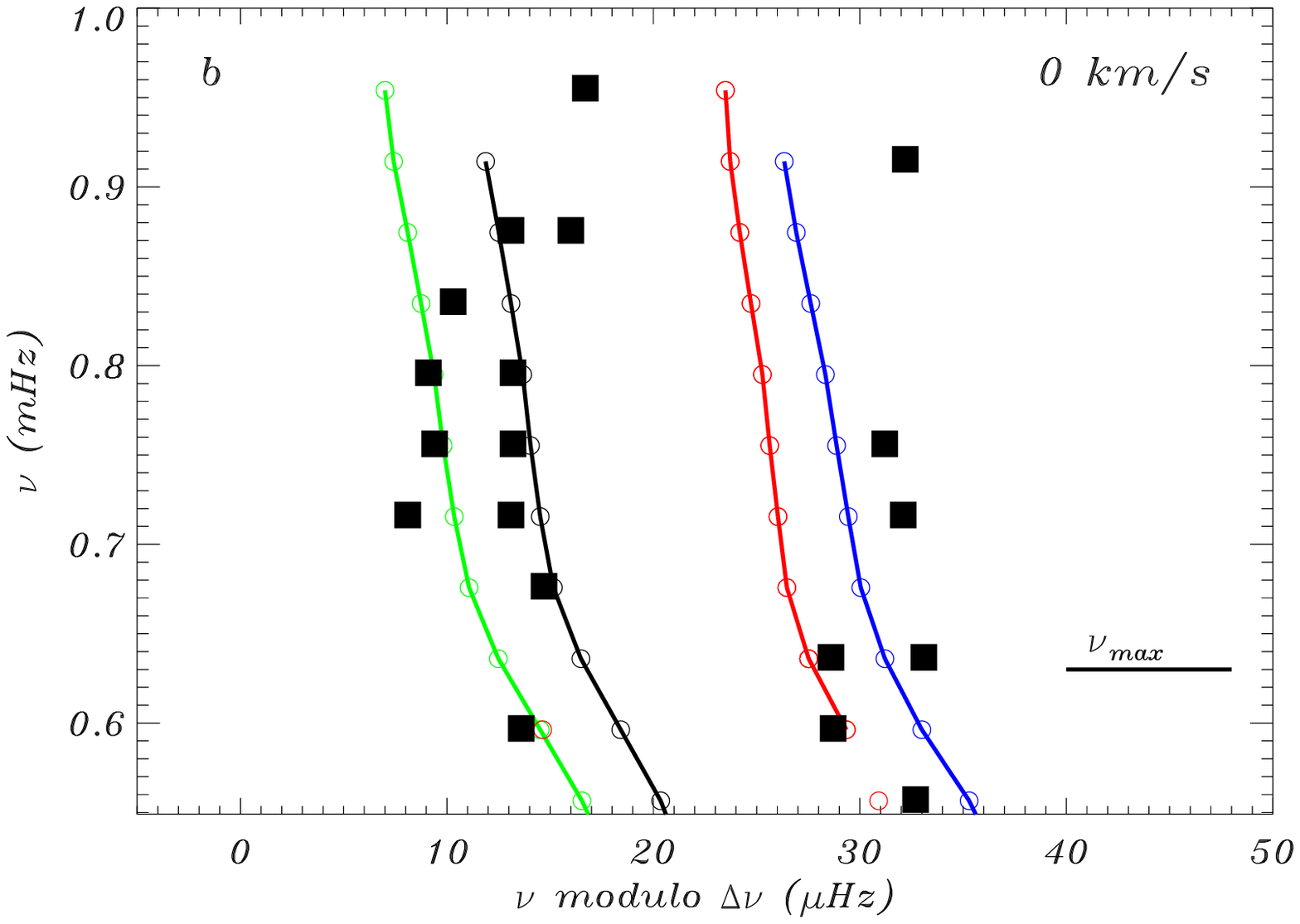} 
    \caption{Illustration of considering rotation effects on echelle diagrams
             (a) for the seismic diagnostics for the solar-like star
             \etaboo\ (oscillation frequencies taken from C05)
             compared with the non-rotating case (b). The observed frequencies are represented by
             black filled squares. The nomenclature used to represent the theoretical 
             frequencies in previous figures is also adopted here. For the sake of clarity,
             the $m=0$ as well as the radial modes are connected with lines. Note the
             presence of mixed modes at low frequency (near $0.6\,\mHz$). For the simplicity
             of the whole figure, these modes are not connected by lines. The uncertainty
             of the observed frequencies, $1\,\muHz$ approximately, were roughly estimated from
             those of the small spacings given in C05 (more details in the text). This is
             approximately the size of the symbols. Stars,
             triangles, and inverted triangles represent modes with $|m|=1$, 2, and 3,
             respectively.}
\label{fig:ed_etaboo}
\end{figure*}

\subsection{High frequency range and surface effects \label{ssec:highfrequency}}

Second-order effects of rotation increase with the frequency and are therefore largest in the high
frequency domain (above $\nu_{max}$) where frequencies are sensitive to various surface effects
\citep[see][ for a review on these effects]{GoupilDupret07}. For instance, \citet{Straka06} showed
that turbulence causes variations in the upper part (highest-frequencies) of EDs. In that work, a
comparison of classical models without turbulence with models including 3D simulations of
turbulence reveals that differences in frequency for radial modes of about $5\,\muHz$ (from $\numax$
to higher frequencies) between the two types of models can exist (see EDs of Figs.~4 and 5 in Straka
et al.). These orders of magnitude are comparable to those we find here for models with rotational
velocities of about $16\,\kms$  in the equivalent frequency domain. 

Similarly, it has been shown \citep{Theado05, CastroVauclair06} that the effects on the model
frequencies with and without including such diffusion processes are different for radial modes than
for non-radial modes thereby affecting the small spacings. Such effects can reach a few $\,\muHz$
for a model of $1.30\,\msun$. This means that, rotational second-order effects shown in the EDs of
Figure~\ref{fig:edrurd} can be larger than those coming from microscopic diffusion of He even for
very small rotational velocities. Rotation can thus be considered as another contribution to such
surface effects, like turbulence. Note however that rotation effects extend to  smaller frequencies
for higher rotation rates.

Scaled small frequency spacings like $r_{1,0}$ (see Section~\ref{ssec:fspacings}) have been shown to
be rather insensitive to surface effects, and therefore can probe properties of the inner part, like
the convective core overshoot \citep{RoxVor03,RoxVor04}. This is no longer true when rotation is
responsible for the 'surface effects'. Indeed we find a similar behavior of both spacings $r_{1,0}$
and $\dnu_{1,0}$ (Figs.~\ref{fig:spacings}c and\ref{fig:spacings}b, respectively). The reason 
is that the effects of rotation on EDs are mainly explained by the variations in the small spacings.
These variations become significant when the rotational velocity and oscillation frequency increase
(see Figs.~\ref{ssec:fspacings}a \& \ref{ssec:fspacings}b,) but they are not compensated by the
effects on large spacings which remain quite small. This is confirmed with a calculation of the
$r_{1,0}$ removing the effect of the stellar distortion by the centrifugal force from the
oscillations frequencies, i.e. we neglect the second-order terms in the perturbation treatment.
Then, all the curves in Figure~\ref{fig:spacings}c collapse (Figure~\ref{fig:spacings}d).

These results imply that even at small stellar rotational velocities (from 10-$15\,\kms$), the
scaled $r_{\ell,n}$ spacings are sensitive to the stellar distortion, therefore care must be taken
when using  them as indicators for probing the stellar interior. It is possible to find
appropriate combinations of frequencies which enable to remove 'pollution' due to second-order
effects of rotation in $r_{1,0}$ and $\dnu_{1,0}$ 
\citep{DziembowskiGoupil98, Goupil04,LochardThesis}

\subsection{An illustration: the solar-like star \etaboo \label{ssec:etaboo}}

In order to illustrate how rotation affects the asteroseismic diagnostics of real data, we
selected the well-known solar-like star \etaboo\ (\object{HD HR5235}) which has a measured $vsini$
of $12.8\,\kms$ \citep{BouchyCarrier02}. This bright sub-giant G0 star has been the subject of many
observational and interpretation works in the literature. One of the most recent works on
theoretical interpretation of this star was published by \citet{Carrier05etaboo} (hereafter C05). In
that work, a detailed asteroseismic modeling of the star, including rotation and atomic diffusion,
was performed. For illustrative purposes, we computed stellar models which are approximately 
similar to some of the representative models listed in Table~5 of the aforementioned paper.  The
physical parameters and the observed oscillation frequencies of the star were taken from C05.

As shown in the previous sections, shellular rotation and uniform rotation profiles yield almost
identical results at the scale of EDs. Thus, in order to simplify the procedure, we adopt, for the
present illustration, the hypothesis of uniform rotation. For the convection description, the
mixing-length formulation is used, assuming an overshoot of $0.2$. The physical parameters
and the observed oscillation frequencies of the star were also taken from C05. 

We searched for models matching the observed frequencies and the large spacing estimated by
C05 ($\alsep=39.9\,\muHz$) with the help of EDs. To do so, stellar models with rotational
velocities ranging from $12\,\kms$ ($i\rightarrow90\deg$) and $48\,\kms$ ($i\rightarrow14\deg$)
were computed. This range of rotational velocities ensures the use of a perturbative approach
for the oscillation computations (see discussion of the perturbative approach in
Section~\ref{sec:comppoly}).

Assuming that most detected modes are $m=0$ modes, the best diagnostics for the large separation
($\alsep=39.78\,\muHz$) were obtained for a model, ${\cal M}_{\eta}$, with similar characteristics
to model M3 of C05, i.e., a mass of $1.70\,\msun$ with radius $R=2.79\,\rsun$, and an age of
2.25\,Gyr ($\teff=6014\,\kelvin$, $L=9.14\,\lsun$), and surface rotational velocity of $30\,\kms$
(which implies an inclination angle of $i\sim25^\circ$). A value $\alpha=1.75$ is taken for the MLT 
mixing length parameter. The chemical composition is fixed at  $X=0.7, Z=0.03$.  Additional
characteristics of this model are listed in Table~\ref{tab:models}.

The resulting ED for ${\cal M}_{\eta}$ fits reasonably well the $m=0$ modes with the observations
(see Figure~\ref{fig:ed_etaboo}a) whereas for a lower rotation velocity set at 21 $\,\kms$, the
match is poorer and the stellar model would not be considered as a good match in absence of rotation
(Figure~\ref{fig:ed_etaboo}b), particularly for the $\ell=1,m=0$ modes. Even when rotation is
included, the mode identification is not unique. Note that the (1,-1), (3,-3), and (3,-2) ridges are
very close to $(1,0)$, which is also the case of $(1,+1)$ with $(3,0)$. Similarly, we found that the
ridges $(2,-2)$ and $(2,-1)$ are almost coincident with $(0,0)$, and the $(2,+2)$ with $(2,0)$. 
Some of the high-frequency modes ($\nunl\geq0.85\,\mHz$) are better fit with $m\neq0$ ridges. For
the selected model, these modes are hardly fit when rotation effects are not included in the
modeling, as shown in panel b of Figure~\ref{fig:ed_etaboo}. In any case, we are aware that the
presence of all $m$ components makes the ED more dense. This increases the probability of a better
match and non-uniqueness of the solution, in particular when no independent mode identification
information is available. Model fine tuning should thus consider variations of both the amount and
the shape of the rotation profile.

Quantitatively, the effects of rotation found reach $5\,\muHz$ approximately in certain cases (e.g.
for $\ell=1$ modes) when considering only $m=0$ modes, but this is largely increased when fitting
$m\neq0$ ridges. In general, these effects are much larger than the observational
uncertainties. Since no information on individual frequency uncertainties are given in C05, the
observational uncertainties shown in Figure~\ref{fig:ed_etaboo} correspond to the uncertainty of
the observational $\dnu_{0,2}$ small spacings given in that work, i.e. around $1\,\muHz$
($\dnu_{0,2}$ corresponds to the mean separation between the $\ell=0$ and 2 ridges). Note that
it is possible that the present solution including rotation may not be unique. For instance, small
variations around the physical characteristics of ${\cal M}_{\eta}$ (mass, effective temperature,
gravity, angular momentum distribution, etc.) may yield similar fits. It can be shown that
variations in the physical characteristics of ${\cal M}_{\eta}$ must be small in order fit the
observations as shown in Figure~\ref{fig:ed_etaboo}. This is coherent with an echelle diagram of an
evolved star (in the main sequence) as it is predicted by model ${\cal M}_{\eta}$. This explains
also that small variations of the modulo, around $0.01\,\muHz$ affect significantly to the accuracy
of the fit.

Depending on the star characteristics, the impact on precise seismic diagnostics and and thereby on
accurate modeling varies. For very small rotational velocities, i.e. $vsini\leq5\,\kms$, rotation
effects can be neglected. However, for higher $\vsini$, the modeling should be performed taking the
second-order (distortion) effects into account. Special caution must be taken when no additional
information about the angle of inclination of the star is available (and/or the rotational
splitting), otherwise the accuracy of the modeling may be compromised significantly. This, combined
with additional information coming from spectroscopy and/or multicolor photometry, may be helpful
for the identification of modes in solar-like stars (Su\'arez et al., work in progress using \corot\
data). 

This example illustrates well that for solar like oscillating, low and intermediate mass main
sequence stars with rotation rates in the upper part of the observed rotation rate range, the
modeling should be performed taking full second-order (distortion) effects into account. Hence, 
special caution must be taken when no additional information about the angle of inclination of the
star is available (and/or the rotational splitting), otherwise the accuracy of the modeling may be
significantly compromised if the rotation of the star happens to be large despite a small $\vsini$.


\begin{acknowledgements}
   JCS acknowledges support from the "Instituto de Astrof\'{\i}sica de Andaluc\'{\i}a (CSIC)" by an
   "Excellence Project post-doctoral fellowship" financed by the Spanish "Conjerer\'{\i}a de
   Innovaci\'on, Ciencia y Empresa de la Junta de Andaluc\'{\i}a" under proyect "FQM4156-2008".
   JCS also acknowledges support by the Spanish "Plan Nacional del Espacio" under project
   ESP2007-65480-C02-01, and financial support from CNES. DRR acknowledges support from the CNES
   through a post-doctoral fellowship.
\end{acknowledgements}


\end{document}